\documentclass[preprint,12pt]{elsarticle}

\usepackage[english]{babel}
\usepackage[x11names,dvipsnames,table,xcdraw,svgnames]{xcolor}
\usepackage{amsmath,amsfonts,amssymb,bm}
\usepackage[colorlinks,linktocpage]{hyperref}
\usepackage[ruled,vlined,english, algosection]{algorithm2e}
\usepackage{enumitem}
\usepackage{subcaption}
\usepackage{tikz}
\usepackage{tikz-layers}
\usepackage{pgfplots}
\pgfplotsset{compat = newest}
\usetikzlibrary{patterns,decorations.pathmorphing,positioning}
\usetikzlibrary{shapes.geometric}
\usetikzlibrary{shapes}
\usetikzlibrary{calc,angles,quotes}
\usetikzlibrary{trees}
\usetikzlibrary{mindmap}
\usetikzlibrary{pgfplots.groupplots}
\usetikzlibrary{arrows.meta}
\usetikzlibrary{intersections, pgfplots.fillbetween}
\usetikzlibrary{plotmarks}
\usepackage{multirow}
\usepackage{booktabs}
\usepackage{nicefrac}
\usepackage{floatrow}
\usepackage{etoolbox}
\usepackage{array}
\usepackage{natbib}
\usepackage{adjustbox}
\usepackage{svg}
\svgpath{inkscape=inkscape --export-type=pdf, svgpath=images/}

\hypersetup{
    colorlinks=true,
    citecolor=Green,
    filecolor=black,
    linkcolor=blue,
    urlcolor=MidnightBlue,
    pdftitle={Statistical inference of upstream turbulence intensity for
the numerical analysis of the flow around a bluff body with massive separation}
}

\newcommand \pd[2] { \frac{\partial {#1} } {\partial {#2}} }

\newcommand{\Autorefs}[1]{%
  \begingroup%
    \def\chapterautorefname{Chapters}%
    \def\figureautorefname{Figures}%
    \def\sectionautorefname{Sections}%
    \def\subsectionautorefname{Subsections}%
    \def\subsubsectionautorefname{Subsubsections}%
    \def\paragraphautorefname{Paragraphs}%
    \def\tableautorefname{Tables}%
    \def\equationautorefname{Equations}%
    \def\algorithmautorefname{Algorithms}%
    \def\lineautorefname{Lines}%
  \autoref{#1}%
  \endgroup%
}

\journal{Elsevier}

\begin{document}

\begin{frontmatter}


\title{Statistical inference of upstream turbulence intensity for the flow around a bluff body with massive separation}

\author[inst1]{Tom Moussie}
\affiliation[inst1]{organization={Univ. Lille, CNRS, ONERA, Arts et Métiers ParisTech, Centrale Lille, UMR 9014- LMFL- Laboratoire de Mécanique des fluides de Lille},
            city={Kampé de Feriet},
            postcode={F-59000}, 
            state={Lille},
            country={France}}

\author[inst1,inst2]{Paolo Errante}
\author[inst1]{Marcello Meldi}

\affiliation[inst2]{organization={L2EP, Laboratoire d'Electrotechnique et d'Electronique de Puissance de Lille},
            addressline={Avenue Henri Poincaré}, 
            city={Bâtiment ESPRIT},
            postcode={59655}, 
            state={Lille},
            country={France}}

\begin{abstract}
The Benchmarck on the Aerodynamics of a Rectangular 5:1 Cylinder is studied using a data-driven technique which bridges numerical simulation and available experimental results. Because of intrinsic features of the tools used for investigation, in particular in terms of set-up and boundary conditions, significant discrepancies have been observed in the literature when comparing experimental and numerical results. An approach based on the Ensemble Kalman Filter (EnKF) is here used to optimize a synthetic turbulent inlet used as boundary condition in the numerical calculation, in order to reduce the discrepancy with the available experiments. The data-driven method successfully optimizes the boundary condition features, which produce a significant improvement of the accuracy in the prediction of the flow. These finding open perspectives of application towards the analysis of realistic cases, where boundary conditions are complex and usually unknown.
\end{abstract}


\begin{highlights}
\item A data-driven approach based on the Ensemble Kalman Filter (EnKF) is used to augment a scale resolving, hybrid RANS-LES solver using experimental data.
\item The process of DA augmentation, which is based on the optimization of free parameters of the inlet condition, significantly improved the global flow prediction. These finding open perspectives of application towards the analysis of realistic cases, where boundary conditions are complex and usually unknown.
\end{highlights}


\end{frontmatter}


\section{Introduction}
\label{sec:sample1}

Among the open challenges in wind engineering applications for urban settings, the accurate prediction of unsteady features and higher order statistical moments of the flow field represents a key element for technological advancement. Detailed information about the instantaneous organization of the flow can be essential to predict and control the emergence of extreme events.
In addition, the accurate representation of instantaneous features of the flow is a key element to obtain precise estimation of the higher-order statistical moments of the physical phenomena at play. Such statistical moments, which are not well described by stationary closures such as Reynolds-Averaged Navier--Stokes (RANS) models \cite{pope_turbulent_2000,Wilcox2006_DCW}, play an important role in the mechanical stress of urban structures and have to be taken into account in the concept and design phases of production. For example, the second order moment (variance) of the pressure field is essential to measure the surface stress affecting buildings \cite{buildings9030063} and is can also be tied to dangerous aeroelastic phenomena which can affect slender structures \cite{MATSUMOTO1993873}. Experiments in wind tunnels can measure such instantaneous features of the flow. However, the positioning of velocity sensors, such as hot wires, and pressure taps can be precluded in sensitive physical region of the model and systematic usage in realistic applications is not realistically attainable. On the other hand, the development of new computational architectures and the availability of the resources of supercomputing centers provide reliable tools to investigate such problems using numerical strategies based on Computational Fluid Dynamics (CFD). In particular, reduced-order approaches able to resolve the large scales of the flow, such as wall-modeled Large Eddy Simulation (LES) \cite{pope_turbulent_2000,Sagaut2006_springer} or hybrid RANS-LES \cite{pope_turbulent_2000,Sagaut2006_springer,shur_hybrid_2008}, show promising features for the representation of unstationary flows. In fact, these techniques can naturally provide a complete volume description of the flow, unlike most experimental techniques. In addition, they capture the unstationary, three-dimensional features of high-Reynolds regimes while requiring reduced computational resources when compared with Direct Numerical Simulation (DNS). The main drawback of such reduced-order techniques is the need to introduce turbulence models / wall functions in the discretized numerical problem. These models are extremely sensitive to their parametric description and can introduce a bias in the numerical results.

While both experiments and numerical approaches show advantageous features and drawbacks for the analysis of unsteady flows, both families of tools have to face an additional challenge to provide an accurate prediction. This difficulty lies in the lack of knowledge about initial and boundary conditions. High-Reynolds regimes such as the ones observed in urban settings are extremely sensitive to minimal changes in upstream conditions, which can govern the instantaneous evolution and be responsible for the emergence of extreme events. Such realistic features are difficult to be replicated in an isolated environment such as a wind tunnel for experimental analysis. Numerical simulation has the potential to take into account these upstream flow details, even if non-linear interactions of error sources associated with discretization, modelling and boundary conditions must be well assessed to obtain an accurate prediction.

In the present work, the capabilities of CFD scale resolving methods are investigated for the analysis of a well known flow configuration, the Benchmark on the {Aerodynamics} of a {Rectangular} 5:1 {Cylinder} (BARC) \cite{bruno_benchmark_2014} for $Re= 2 \times 10^5$. This test case is challenging to be analyzed via scale resolving CFD, because of the interaction of numerous physical aspects that must be captured and are which difficult to model. For this reason, a high sensitivity of CFD analyses to variations in the numerical set-up is observed \cite{Mariotti2017,Rocchio2020}. Within this framework, the CFD hybrid method known as Delayed Detached Eddy Simulation (DDES) \cite{Gritskevich_IDDES} will be augmented in order to take into account upstream instantaneous features of the flow, such as the upstream turbulent intensity \cite{MANNINI201742,Pasqualetto2021_ws}. To do so, the parametric description of a synthetic turbulence generator used at the CFD inlet boundary condition will be calibrated to reduce the discrepancy with experimental data available from CSTB Nantes \cite{CSTB_experiments}. This optimization problem will be performed using the Data Assimilation technique known as Ensemble Kalman Filter (EnKF) \cite{bocquet_data_2016}.

The manuscript is organized as follows. In \autoref{sec:BARC}, an extensive discussion of the test case is performed and the experimental data available for the present analysis is described. In \autoref{sec:CFDnum}, the CFD solvers used in the present analysis are described and the numerical set-up of the test case is discussed. In \autoref{sec:DAEnKF}, the data-driven technique chosen for this analysis, the EnKF, is detailed. In \autoref{sec:assesment}, the sensitivity of the numerical simulations to different parameters, including the behavior of the inlet condition, is investigated. In \autoref{sec:DA-results}, the data-driven technique is used to optimize the inlet boundary conditions. The results are compared with preliminary CFD runs and the experiments available. In \autoref{sec:conclusions}, conclusions are drawn and future perspectives are discussed.

\section{Test case of investigation: the Benchmark on the Aerodynamics of a Rectangular Cylinder (BARC)} \label{sec:BARC}
The Benchmark on the Aerodynamics of a Rectangular Cylinder with chord-to-depth ratio equal to 5 (BARC) \cite{bruno_benchmark_2014} which started officially in 2008, has the goal to analyze the aerodynamics of a rectangular bluff body. As mentioned by Bartoli et al. \citep{Bartoli2009BARCAB}, the benchmark scope is to contribute to the analysis of separated and turbulent flow around a fixed rectangular cylinder, to provide useful information for civil engineering applications, since this kind of geometry is often encountered in urban areas (e.g. long span bridges decks or high-rise buildings). Since then, this configuration has been extensively studied both experimentally and numerically \cite{le_spanwise_2009,cimarelli_direct_2018,rocchio_flow_2020}. A sketch of the configuration is given in \autoref{fig:BARC_test_case_BCs} together with the computational domain used in this study.

\begin{figure}[h!]
    \centering
    \includegraphics[width=.85\textwidth]{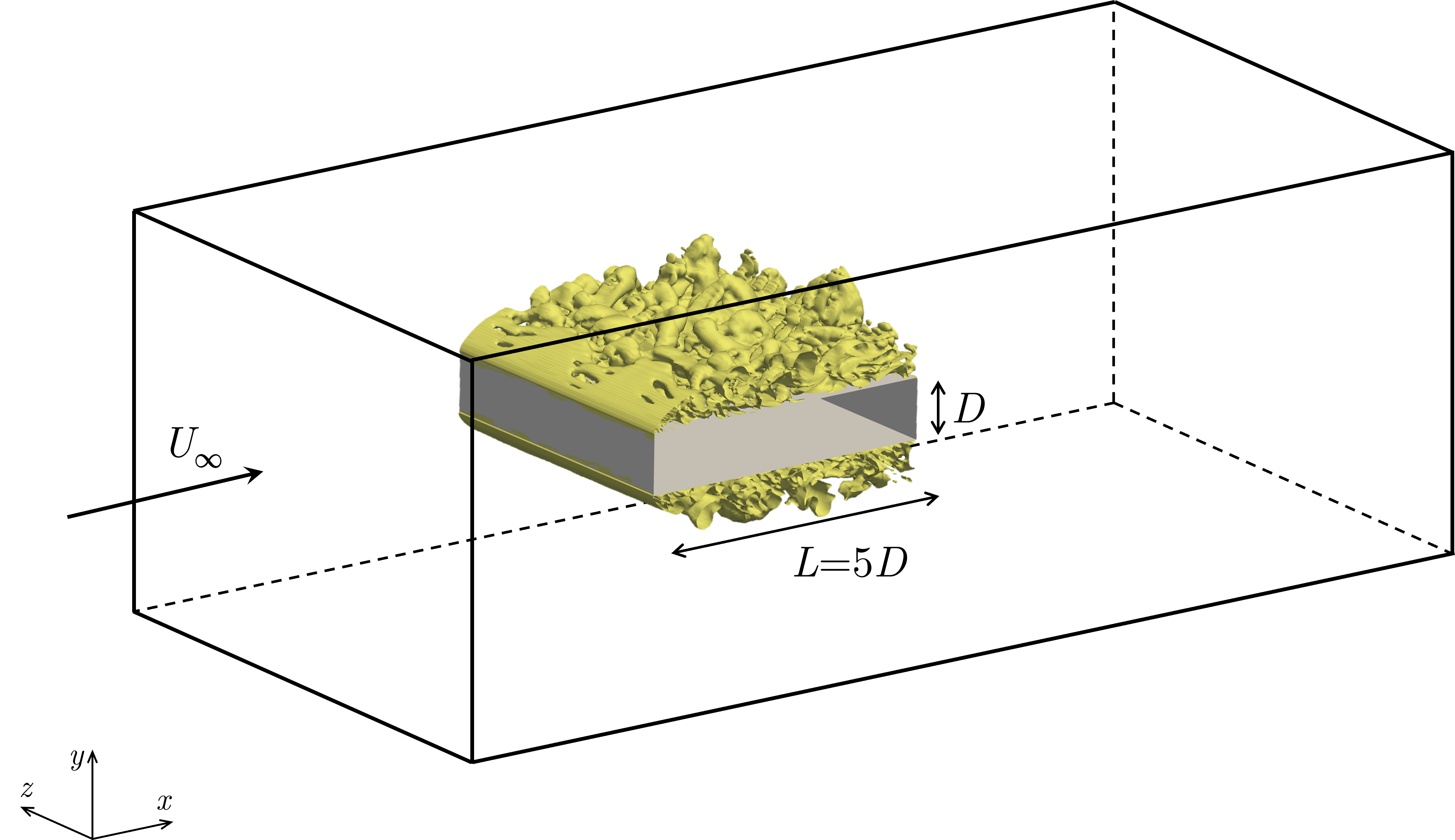}
    \caption{Qualitative representation of the flow around the rectangular cylinder. The $\lambda_2$ criterion is shown to highlight the unstationary features of the flow around the immersed body.}
    \label{fig:enter-label}
\end{figure}

Despite its simple appearance, this configuration involves complex physical processes such as boundary layer separation, flow reattachment, recirculation zones, and Von Kármán streets, which affect the propagation of acoustic waves and the flow's structural organization, as shown for instance in \citet{le_spanwise_2009} and \citet{ricciardelli_sectional_2008}. Therefore, the complete representation of these mechanisms and their interactions is crucial to obtain an accurate prediction for similar cases of industrial and urban interests. One major challenge in studying this flow configuration via numerical methods is its sensitivity to the geometric characteristics and boundary conditions, particularly the velocity field imposed at the inlet. This is especially true for scale-resolving approaches like LES and hybrid RANS-LES, which are now widely used for analyzing bluff bodies \cite{Minguez_pof2008,HESSE2021104901,HAFFNER2022_ijhff}.

\subsection{Review of the analyses in the literature}
 
Numerical analyses of the BARC have mainly focused on the analysis of the sensitivity of this test case to variations in the set-up of the problem, in particular for geometric characteristics and features of turbulence modeling \cite{Bruno2014BenchmarkOT}. For the latter, unsteady RANS (URANS), LES and hybrid turbulence models have been tested by several authors. DNS for relatively low Reynolds numbers have also been proposed in the literature \cite{Cimarelli2018_jweia,Chiarini2021_ftc} which exhibit significant discrepancies in the statistical quantities investigated, confirming the difficulty for numerical approaches to capture the leading physical dynamics for this test case. URANS studies, such as the one by \citet{MANNINI20101609}, indicate a strong sensitivity to the choice of the turbulence model. For LES, \citet{Timilsina2015BenchmarkOT} shown that subgrid-scale (SGS) modeling can achieve satisfying predictions on drag coefficients and with a relatively small impact attributed to the closure used, while \citet{Rocchio2020} find that variations in the parametric choices for the SGS model can be dominant. \citet{mariotti_stochastic_2017} performed a stochastic analysis of LES predictions of the flow around the BARC. The analysis is repeated for two different grid resolutions, which differ in terms of streamwise and spanwise resolution. Conclusions state that for both numerical experiments, the most sensitive quantities are those which show the largest dispersion among the different BARC studies in the literature, such as the lift coefficient, the time-average flow features and the pressure distribution. Also, results obtained with the two grids exhibit remarkable differences, for the recirculation region for instance. Hybrid approaches have been also used by \citet{mannini_numerical_2011} which highlight the importance of numerical dissipation introduced by the numerical schemes. It has been concluded that an excessive amount of dissipation damps out the turbulent structures directly resolved by the grid used, which are essential to capture the emergence and interactions of the numerous physical aspects which characterize this flow.

The analyses in the literature dealing with the sensitivity of the flow to geometric variations have mainly focused on two aspects. The first one deals with the alignment of the rectangular cylinder to the direction of the upstream flow. \citet{Patruno2016_jweia} investigated the statistical features of the flow with variations in the angle of attack. The analysis included large variations of the angle as well as minor variations which could be associated with uncertainty in the set-up of experiments. The second main topic of investigation deals with the shape of the edges of the BARC, which traditionally are considered to be sharp. The study by \citet{rocchio_flow_2020}, which consisted of a comparison of LES with and without edge rounding on the leading edge, indicated a significant discrepancy of the results. In particular, the length of the recirculation region is strictly connected to this geometric feature.

The previous discussion highlights the very high sensitivity of the BARC test case to several parameters governing the set-up. This parameters can be optimized via data-driven strategies, which are now extensively used in fluid mechanics applications \cite{annurev-fluid-010719-060214}. One recent example of application for the BARC is the usage of multigrid sequential data assimilation, which have been used to calibrate SGS models for LES by \citet{moldovan_optimized_2022}. The results presented for the statistical moments of the velocity and pressure flow field, which were obtained for the BARC with $Re = 4 \times 10^4$  show that data assimilation techniques based on the Ensemble Kalman Filter are able to improve the predictive features of the CFD solver for reduced grid resolution. In addition, it has been observed that, despite the sparse and asymmetric distribution of observation adopted in the data-driven process, the data augmented results exhibit symmetric statistics and improved accuracy far from the sensor location. 

Experimental studies in wind tunnels for the BARC \cite{bruno_benchmark_2014,le_spanwise_2009, ricciardelli_sectional_2008,haan_anatomy_2009} also investigate the sensitivity of the flow to several parameters such as the Reynolds numbers, the shape / inclination of the rectangular cylinder and the turbulence intensity of the flow upstream. While important variations of the physical quantities are observed in experiments as well, the discrepancy with numerical simulation is important and significantly larger than the uncertainty in the set-up \cite{Bruno2014BenchmarkOT}. Therefore, comparison of results from different tools does not shed a light on the governing mechanisms driving interactions between the instantaneous / statistical features of the flow and the aerodynamic forces at play.
A key question about the sensitivity of the BARC to variations in the set-up is associated with the upstream conditions of the flow. Experiments can naturally take into account this aspect, which is usually measured with the empty wind tunnel. However, one can argue that once the rectangular cylinder is installed in the test tunnel, features of the upstream flow field are also affected, given the subsonic nature of the flow. With numerical simulations, features can be exactly imposed at the inlet, but they usually are difficult to be exactly estimated. \citet{Mariotti2016} investigated the sensitivity of the results to variations in inlet conditions for RANS models. While their results indicate that the sensitivity was weak, the scope of their analysis was limited to RANS modeling and the statistical physical quantities associated. One could argue that inlet conditions for scale resolving simulations, which can account for instantaneous features of the flow, have the potential to govern the organization of the flow and to provide a more reliable picture of the sensitivity of the BARC to such aspect.   

\subsection{Reference experiments for the present work}

The \textit{Centre Scientifique et Technique du Bâtiment} (CSTB) in Nantes, France, has recently open a new benchmarch for the BARC \cite{CSTB_experiments}. The main difference is the Reynolds number of investigation, which is higher than the previous studies summarized by \citet{Bruno2014BenchmarkOT}. The researchers in CSTB have performed an experimental campaign in their \textit{Jules Verne} wind tunnel, analyzing the flow features with the following variations of parameters:
\begin{itemize}
\item Reynolds number $2 \times 10^5$, $3.3 \times 10^5$
\item Three configurations of the front edges: straight, chamfered and curved. 
\item Angle of attack of the rectangular cylinder: $0^\circ$, $\pm 2^\circ$, $\pm 5^\circ$ degrees
\item Turbulence intensity for the upstream flow $\approx 1\,\%$, $\approx 3.5\,\%$
\end{itemize}
The configuration considered for the present analysis is the one for $Re =2 \times 10^5$, sharp front edges, $0^\circ$ angle of attack and $T_u \approx 3.5\,\%$ inlet turbulence intensity. In this case, the rectangular cylinder with height $D = 0.2$\,m and length $L = 5D$, is positioned in the middle of the wind tunnel whose cross section is $5\, \text{m} \times 6\, \text{m}$ and its total length is $12$\,m. The mean flow is aligned with the streamwise direction $x$ and it is described by a bulk velocity of $U_\infty = 15\,\text{ms}^{-1}$.

Measurements for the velocity the pressure fields are performed. 

\begin{figure}[h!]

\begin{tabular}{cc}
    \includegraphics[width=0.49\textwidth]{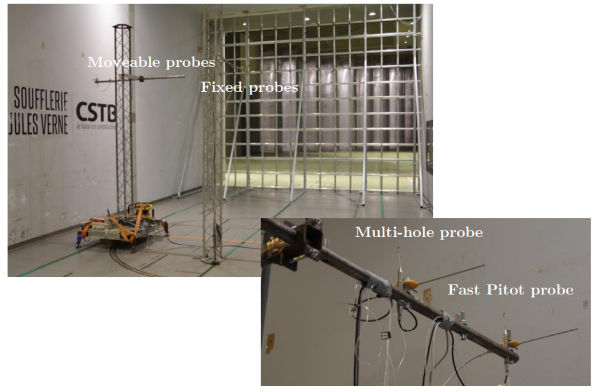} & \includegraphics[width=0.49\textwidth]{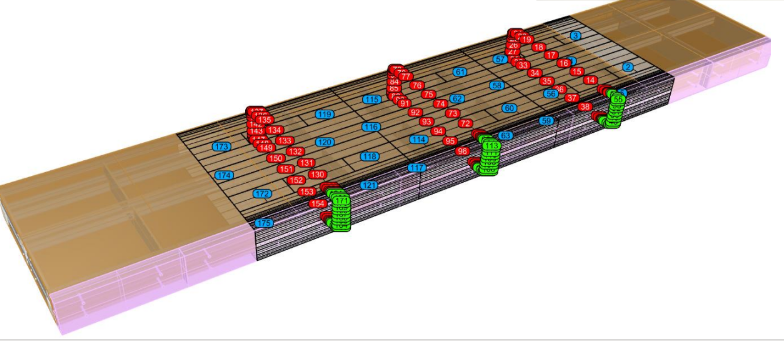} 
\end{tabular}
\caption{Experimental mock up, wind tunnel and sensor positioning for the CSTB campaign \cite{CSTB_experiments}. \label{fig:experiment_measurements}}
\end{figure}

For the latter, 150 sensors on the cylinder's surface are used to obtain time-resolved pressure samples at a frequency of 200\,Hz. The velocity field is investigated using a large-scale Particle Tracking Velocimetry (PTV) over two planes. The first one, which is normal to the spanwise direction and is of size $\Delta_x \times \Delta_y = \left(2L \times L \right)$, is used to obtain time-resolved measurement of the velocity components $u_x$ and $u_y$. The second plane, located at $x=-0.45L$, which is normal to the streamwise direction and is of size $\Delta_y \times \Delta_z = \left(3D \times 3D \right)$, provides measurements for the velocity components $u_y$ and $u_z$. The frequency of acquision of the PTV is equal to 6.6\,kHz. The size of the first plane presented is large enough to sample the flow under the rectangular cylinder as well as part of the wake.
In addition to the measurements presented for the flow features, the time history of the lift and drag coefficients is also provided.

\section{Numerical tools}
\label{sec:CFDnum}

The numerical algorithms used in the present work are now introduced. 

\subsection{Dynamic equations and numerical solvers} 
\label{sec3:DDES}

The numerical solvers used in this work rely on Finite-Volume discretization of the Navier--Stokes equations for incompressible flows and Newtonian fluids. As previously discussed in the Introduction, the analyses will be performed using coarse grained approaches based on turbulence / subgrid-scale modeling. Therefore, considering a reduced-order operator $\tilde{\cdot}$ \, ,which could represent a RANS average or an LES filtering, the dynamic equations can be written as:
\begin{align}
    \nabla \cdot \mathbf{\tilde{u}} &= 0 \label{eq:mass_conservation} \\
    \frac{\partial \mathbf{\tilde{u}}}{\partial t} + \left( \mathbf{\tilde{u}} \cdot \nabla \right) \mathbf{\tilde{u}} &= - \nabla \tilde{p} + \nu \nabla^2 \mathbf{\tilde{u}} - \nabla \cdot {\bm{\tau}_t} \label{eq:momentum_equation}
\end{align}

were $\mathbf{u}$ is the velocity field, $p$ is the pressure (normalized over the density $\rho$), $\nu$ is the kinematic viscosity and $\bm{\tau}_t$ is the tensor representing the effects of the turbulence closure. This term would be Reynolds stress tensor in the case of RANS approaches, while it would be the subgrid stress tensor for LES. For eddy-viscosity models \cite{pope_turbulent_2000,Sagaut2006_springer}, the components of this tensor are determined by the spatial gradients of the resolved velocity field and by the quantity $\nu_t$, which is referred to as turbulent viscosity. Relying on the Boussinesq approximation, it is used to take into account the statistical moments of stresses due to turbulence as a diffusive effect. In this case, the components of the tensor ${\bm{\tau}_t}$  are:
\begin{equation}
    {\tau}_{t,ij} = -2 \nu_t \left( \tilde{S}_{ij} - \frac{1}{3} \mathcal{K}\delta_{ij} \right) = - \nu_t \left( \frac{\partial \tilde{u}_i}{\partial x_j} + \frac{\partial \tilde{u}_{j}}{\partial x_i}  \right) + \frac{2}{3} \mathcal{K}\delta_{ij}
\end{equation}
where $\tilde{S}_{ij}$ are the components of the resolved rate of strain tensor, $\mathcal{K}$ is the turbulent kinetic energy and $\delta_{ij}$ is the Kroneker symbol.
Several proposals in the literature  for eddy viscosity models provide different expressions for the quantity $\nu_t$. In the context of this study, the hybrid RANS-LES method known as Delayed Detached Eddy Simulations (DDES) will be used \cite{spalart_new_2006}. This model, which behaves like a RANS closure in the proximity of the wall, and transitions to an LES behavior moving in free stream turbulent regions, does not require the refinement of wall resolved LES and DNS at the body surface. Therefore, it has been selected because it can capture the three dimensional unsteady features of high-Reynolds flows with reasonable computational costs. In DDES, the turbulent viscosity $\nu_t$ is obtained via an algebraic relation with a newly introduced quantity, the viscosity-like variable $\dot{\nu}$.
\begin{align}
    \nu _{t}={\dot  {\nu }}f_{{v1}} \label{eq:SA_turbulent_eddy_viscosity}
\end{align}

with:
\begin{equation}
f_{{v1}}={\frac{\chi^{3}}{\chi^{3}+C_{{v1}}^{3}}} \; , \qquad \chi ={\frac{{\dot{\nu}}}{\nu}}
\end{equation}

$C_{{v1}}$ is a constant to be provided by the user. The variable $\dot{\nu}$ is obtained resolving a transport equation, like in classical one-equation eddy viscosity models:

\begin{align}
\frac{\partial \dot{\nu} }{\partial t} + \left( \mathbf{\tilde{u}} \cdot \nabla \right) \dot{\nu} = \overbrace{\mathcal{P}_{\dot{\nu}}}^{\text{Production}} + \overbrace{ \mathcal{T}_{\dot{\nu}}}^{\text{Diffusion}} - \overbrace{\mathcal{D}_{\dot{\nu}}}^{\text{Destruction}}
\label{eq:SA_viscosity_transport}
\end{align}

where:
\begin{align}
        \label{eq:SA_production}
        \mathcal{P}_{\dot{\nu}} &= C_{b1}[1-f_{t2}]{\dot {S}}{\dot {\nu }}\\ 
        \mathcal{T}_{\dot{\nu}} &= {\frac {1}{\sigma }}\left(\nabla \cdot [(\nu +{\dot {\nu }})\nabla {\dot {\nu }}]+C_{b2}|\nabla {\dot{\nu}}|^{2}\right) \\
        \label{eq:SA_destruction}
        \mathcal{D}_{\dot{\nu}} &= \left[C_{w1}f_{w}-{\frac {C_{b1}}{\kappa ^{2}}}f_{t2}\right]\left({\frac {\dot {\nu }}{\dot{d}}}\right)^{2} 
\end{align}

One can see that \Autorefs{eq:SA_production} to \ref{eq:SA_destruction} are governed by a number of free coefficients ($C_{b1}, \, C_{b2}, \, C_{w1}, \, \sigma$), physical parameters (von Karman constant $\kappa$) and functions ($f_{w}, \, f_{t2}$) which can be chosen or modified by the user. In particular, the terms ${\dot {S}}$ and $\dot{d}$ determine the production and destruction terms of \autoref{eq:SA_viscosity_transport}. ${\dot {S}}$ is determined by a series of algebraic equations, which include new model coefficients to be selected:
\begin{align}
    \begin{split}
        \dot{S} \equiv S+{\frac  {{\dot  {\nu }}}{\kappa ^{2}d^{2}}}f_{{v2}},&\qquad f_{{v2}}=1-{\frac  {\chi }{1+\chi f_{{v1}}}} \\
        S={\sqrt  {2 W _{{ij}} W _{{ij}}}}, & \qquad W _{ij}={\frac{1}{2}} \left( \pd{u_i}{x_j} - \pd{u_j}{x_i} \right) \\
        f_{{t2}}=C_{{t3}}\exp \left(-C_{{t4}}\chi ^{2}\right) , & \qquad f_{w}=g\left[{\frac  {1+C_{{w3}}^{6}}{g^{6}+C_{{w3}}^{6}}}\right]^{{1/6}} \\
        g=r+C_{{w2}}(r^{6}-r),&\qquad r\equiv {\frac  {{\dot  {\nu }}}{{\dot  {S}}\kappa ^{2}d^{2}}} \\ \label{eq:SA_equation_parameters}
    \end{split}
\end{align}

The definition used for the parameter $\dot{d}$, which represents a characteristic length, is intrinsically associated with the hybrid RANS-LES chosen technique. In the context of the DDES model, $\dot{d}$ is defined as:
\begin{align}
\label{eq:d}
    \dot{d} = d- f_\mathrm{d} \max \left( 0,d-C_{\text{DES}}\Delta \right)
\end{align}

Where $d$ is the distance from the closest wall, $C_{\text{DES}} = 0.65$ and $\Delta \equiv \max\left(\Delta x, \Delta y, \Delta z \right)$.
In addition:

\begin{align}
    f_{\mathrm{d}} \equiv 1 - \tanh\left( \left( 8\cdot r_\mathrm{d} \right)^3 \right) \, ; \quad  r_{\mathrm{d}} = \frac{\nu_t + \nu}{\kappa^2 d^2 \cdot \max \left( \sqrt{ \left(\nabla \mathbf{\tilde{u}}\right) \cdot \left(\nabla \mathbf{\tilde{u}} \right) }, 10^{-10} \right)}
    \label{eq:f_d}
\end{align}

The algebraic expression for $\dot{d}$ provided in \autoref{eq:d} can blend the model behavior between full RANS and full LES modes. In particular, a dominant RANS behavior will be obtained in proximity of the wall. Increasing the wall distance $d$, $r_{\mathrm{d}} \to 0$ and therefore $f_{\mathrm{d}} \to 1$. In these conditions, a dominant LES behavior is observed. Moreover, one can see that $\dot{d}$ is not exclusively grid-dependent, as $r_{\mathrm{d}}$ exhibits an explicit dependence to $\mathbf{\tilde{u}}$, $\nu_T$ and $\nu$. This choice alleviates some constraints about grid-dependency of the solution, even if a grid sensitivity analysis is always recommended. For sake of completeness, the value of the model constants are reported in \autoref{tab:SA_constants}.

\begin{table}[h!]
\centering
\begin{tabular}{cccccccccccc} \toprule[1.5pt]
$\sigma$ & $C_{{b1}}$ & $C_{{b2}}$ & $\kappa$ & $C_{{w1}}$ & $C_{{w2}}$ & $C_{{w3}}$ & $C_{{v1}}$ & $C_{{t1}}$ & $C_{{t2}}$ & $ C_{{t3}}$ & $C_{{t4}}$ \\
2/3 & 0.1355 & 0.622 & 0.41 & 3.2391 & 0.3 & 2 & 7.1 & 1 & 2 & 1.1 & 2 \\
\bottomrule[1.5pt]
\end{tabular}
\caption{Values chosen for the model constants for the DDES closure used in the present analysis. These values are the ones provided as default in the implementation of the code.}
\label{tab:SA_constants}
\end{table}

Details about the CFD code are now provided. The simulations are performed using the C++ open-source framework OpenFOAM \cite{greenshields2021}. This library includes a number of solvers that can be used to investigate physical configurations exhibiting different complexities and, because of its versatility, it has been extensively used in industrial research as well as for academic studies \citep{Tabor2010553,Meldi2012_pof,Selma2014241,Constant2017_cf}. In this work, two flow solvers have been used, the Simple \cite{caretto_simple_1973} and Pimple \cite{caretto_simple_1973,ISSA198666} algorithms. Both solvers rely on a recursive procedure, where the velocity field and the pressure gradient are iteratively updated until convergence. The solver SimpleFOAM is used in this analysis for stationary simulations, while the PimpleFOAM solver is used for unstationary calculations. For the unstationary simulations, the time discretization relies on a second-order backward scheme and space discretization is performed using second order schemes. In particular, the advective term is discretized using a native LUST scheme, which combines with a ratio $75\%$-$25\%$ a second-order centered scheme and a second-order upwind scheme. For the stationary RANS simulations, a first-order upwind scheme is used for the advective term.

\subsection{Numerical set-up of the test case}
\label{sec3:setup}

\begin{figure}[h!]
    \centering
    \includegraphics[width=.85\textwidth]{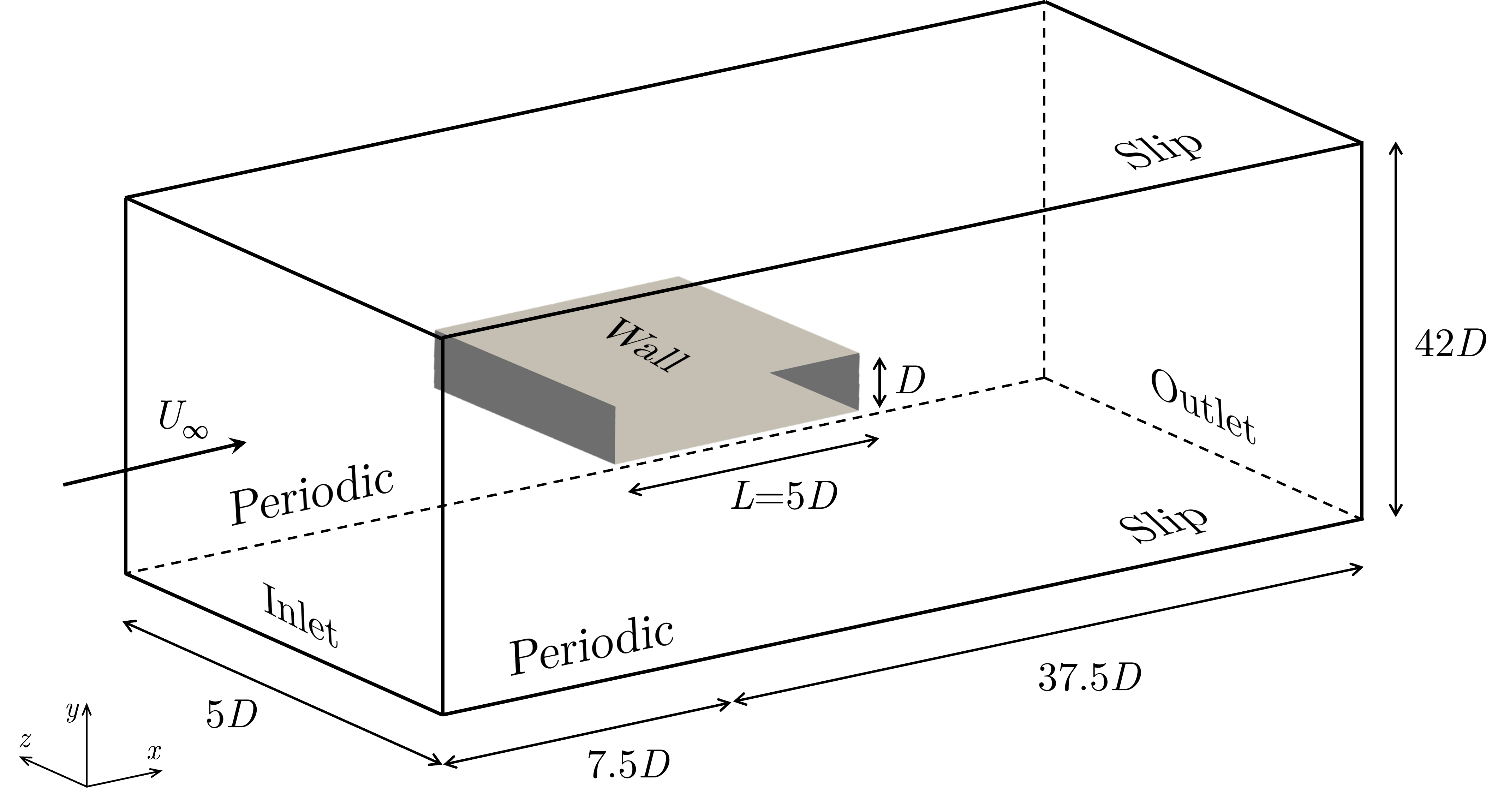}
    \caption{Set-up and boundary conditions for the BARC test case.}
    \label{fig:BARC_test_case_BCs}
\end{figure}

The physical domain of investigation and the boundary conditions applied are shown in \autoref{fig:BARC_test_case_BCs}. 
The axis are set so that $x$ represents the streamwise direction, $y$ is the normal direction and $z$ is the spanwise direction. The resolution of the dynamic system is performed normalizing the physical quantities over the upstream mean velocity $U_\infty$ and the height of the rectangular cylinder $D$. Therefore, the kinematic viscosity for the calculations is set to $\nu = Re^{-1}=(2\cdot10^5)^{-1}$. 

Periodic conditions are imposed on the lateral faces in the $x-y$ planes, while freestream boundary conditions are applied at the top and bottom faces on $x-z$ planes. A mass flow rate conserving boundary condition is applied at the outlet. The parameters of this specific boundary condition, defined as \textit{inletOutlet} in OpenFOAM, have been set so that reverse flow is excluded. For the inlet, two proposals have been considered. The first one is a classical set-up for which the velocity field is uniform and equal to $U_\infty$ while a zero-gradient condition is used for the pressure. The second inlet condition used in the present analysis uses a synthetic turbulence generator based on the works by \citet{poletto_new_2013} and \citet{shur_DFSEM_2018}. This condition is already implemented in OpenFOAM and determines the inlet velocity $\mathbf{\tilde{u}}$ as the sum of the bulk velocity $\mathbf{u}_b = [U_\infty, \, 0, \, 0]$ and a fluctuating velocity $\mathbf{u}^\prime$. 
The methodology is based on the concept of synthetic eddies, which are tied with prescribed velocity fluctuations and are injected on the inlet plane. The eddies are defined by their center and a description of the velocity fluctuations distribution around them. These eddies are randomly generated at the inlet and advected in the physical domain. This process of random generation is constrained by physical features of the velocity field that can be imposed by the user, such as the components of the Reynolds stress tensor or the integral length scale $\ell$. From \citet{shur_DFSEM_2018}, the fluctuating velocity field is computed as follows:
\begin{align}
        \mathbf{u}^\prime(\mathbf{x}) = C_1 \sum_{k=1}^N \mathbf{q} \circ \left(  \mathbf{r}^k \times \bm{\alpha}^k \right) \label{eq:u_i}
\end{align}

where $\mathbf{r}^k = \frac{\bm{x} - \bm{x}^k}{\bm{\sigma} ^k}$ is the normalized distance from the eddy center. $\bm{x}$ is a position vector and $\bm{x}^k$ is the center of the $k$-th eddy. In addition, the vectors $\bm{\sigma}^k$ and $\bm{\alpha}^k$ represents respectively the length scales and the intensity of the $k$-th eddy in each direction. $\mathbf{q}$ is a shape function based on the distance from the eddy centre, whose formulation from \citet{poletto_new_2013} ensure divergence-free velocity field. 
The components of the eddy intensity $\bm{\alpha}$ and length scale $\bm{\sigma}$ are given as following:

\begin{align}
     \alpha^2_{i} = \gamma \frac{\sum_j \left( \nicefrac{\lambda_j}{\sigma_j^2} \right) - 2\nicefrac{\lambda_{i}}{\sigma^2_{i}}}{2C_2} \, ;& \qquad \sigma _i = \max \left(  \delta , c \ell, \, \max\left(\Delta x, \Delta y, \Delta z \right)\right)
    \label{eq:alpha_sigma}
\end{align}

Where $\gamma$ is a random integer which can take the value of $-1$ or $1$ with equal probabilities. The eddy intensity is based on $\lambda$, namely the eigenvalues of the user-provided Reynolds stress tensor. It is worth to notice that formulation of $\alpha _i$ is similar to an ellipsoid having the components of the length scale $\sigma_i$ as semi-axis, and oriented along the local principal reference system to reproduce anisotropy. $C_2$ is a normalization constant that takes into account the magnitude of the shape function integral, whose values are suggested by \citet{shur_DFSEM_2018}.
Assessing the eddy length scale is not a trivial task. Synthetic Turbulence Generation (STG) involves a predefined model energy spectrum which comes from the definition of $\sigma_i$. In the present case, according to \autoref{eq:alpha_sigma}, one has to provide a reasonable estimation on the integral length scale $\ell$ and the domain characteristic length scale $\delta$. $\ell$ is balanced by a constant value $c$, either the Von Kármán or an user-defined constant.

In the context of hybrid RANS--LES modeling, an algebraic expression is often derived to approximate this length scale \cite{guo2023characteristic}, although it may not adequately consider the impact from the flow history and boundary information without explicit modeling the transport equation associated with this length scale. This deficiency can potentially influence the process of emulating synthetic turbulence to match realistic experimental conditions. In the next section, a Data-Assimilation approach is described with the objective to overcome this scarcity of details.

\section{Data Assimilation - Ensemble Kalman Filter}
\label{sec:DAEnKF}

The technique of Data Assimilation (DA) that will be used to infer the parametric behavior of the inlet is now introduced. The Kalman Filter (KF), proposed by \citet{kalman_new_1960}, is a sequential DA tool used to obtain an estimation of the physical variables and / or a set of parameters of a system at a given time. The estimation relies on multiple sources of information, which are characterized by a level of uncertainty supposedly known. The classical version of the KF relies on a set of observations $\boldsymbol{y}$ and a prior state $\boldsymbol{x}$ which is obtained via a linear model $\mathbf{M}$. The updated (\textit{augmented}) state is obtained in an analysis step, where the discrepancy between model prediction and available observation is used to update the former:

\begin{equation}
\boldsymbol{x}^a = \boldsymbol{x}^f + \boldsymbol{K} \left( \boldsymbol{y} - \mathcal{H}(\boldsymbol{x}) \right)
\label{eq:KalmanFilter}
\end{equation}

Here the suffix $f$ refers to the model forecast and the suffix $a$ to the state obtained after the analysis phase. The projection operator $\mathcal{H}$ maps the prediction of the model over the space where the observation is sampled. Usually, $\mathcal{H}$ performs as an interpolator, providing the model solution in the location of the sensors. The update of the physical state is governed by the so called Kalman gain matrix $\boldsymbol{K}$, which is obtained via manipulation of the error covariance matrix $\boldsymbol{P} = \mathbb{E} \left( (\boldsymbol{x} - \mathbb{E}(\boldsymbol{x})) (\boldsymbol{x} - \mathbb{E}(\boldsymbol{x}))^{\top}\right)$. The classical KF formulation is not designed to be used in CFD applications. Navier--Stokes equations include non-linear terms, therefore, the derivation of a linear model $\mathbf{M}$ may imply a significant loss of accuracy. Secondly, the size of the matrix $\boldsymbol{P}$ is proportional to the number of degrees of freedom of the model $n$ i.e. number of mesh elements times number of physical variables considered for CFD. The extensive manipulation of $\boldsymbol{P}$ required to obtain $\boldsymbol{K}$, including a matrix inversion, would demand prohibitive computational resources for realistic CFD applications.

The two critical issues previously mentioned can be mitigated using the Ensemble Kalman Filter (EnKF) first proposed by \citet{evensen_sequential_1994}. Within this context, an ensemble of $n_e$ prior states is advanced in time between consecutive analysis phases using a model $\mathcal{M}$ which can be non-linear. Let us assume to consider an analysis phase at the time step $k$.

A state matrix $\boldsymbol{X}_k \in \mathbb{R}^{n \times n_e}$ is constructed, where each column $i$ corresponds to the physical state of an ensemble member $\boldsymbol{x}_i^f \in \mathbb{R}^{n}$. With the EnKF, the covariance matrix $\boldsymbol{P}$ is not advanced in time anymore and is obtained via a Monte--Carlo approximation, thus:

\begin{align}
    \boldsymbol{P}^f = \bm{\Gamma}_k^f \left( \bm{\Gamma}_k^f \right) ^{\top}
\end{align}

where $\boldsymbol{\Gamma}_k^f \in \mathbb{R}^{n \times n_e}$ is the state anomaly matrix  representing the normalized deviation of the state vectors from their ensemble means. The $i^{\text{th}}$ column of the state anomaly matrix is obtained as: 

\begin{align}
    \boldsymbol{\Gamma}_{i,k}^f = \frac{\boldsymbol{x}^f_{i,k} - \bar{\boldsymbol{x}}^f_{k}}{\sqrt{n_e - 1}} \, , \qquad \bar{\boldsymbol{x}}^f_{k} = \frac{\sum_{i=1}^{n_e} \boldsymbol{x}_{i,k} ^f}{n_e} 
    \label{eq:anomaly_gamma}
\end{align}

To obtain a well-posed mathematical and numerical problem, an ensemble of $n_o$ observations is obtained through the perturbation of the observation vector available at the time step $k$, $\boldsymbol{y}_k \in \mathbb{R}^{n_o}$. The result of this perturbation is an observation matrix $\boldsymbol{Y}_k \in \mathbb{R}^{n_o \times n_e}$. The $n_e$ columns of the observation matrix are obtained with $\boldsymbol{y}_{i,k} = \boldsymbol{y}_k + \bm{\varepsilon}_i$ for $i \in [1, n_e]$. The added random noise $\bm{\varepsilon}_i$ is described as a Gaussian probability function $\bm{\varepsilon}_i \sim \mathcal{N}\left( 0, \boldsymbol{\varsigma}_k \right)$, where $\boldsymbol{\varsigma}_k \in \mathbb{R}^{n_o \times n_o}$ is the observation covariance matrix. It is worth to underline that $\boldsymbol{\varsigma}_k$ should be constructed from the experimental uncertainties, hence, the use of Gaussian noise to reproduce experimental error is one of the assumption made due to scarcity of information. However, when this information is not available, the usage of a Gaussian perturbation is helpful to obtain a robust performance of the global algorithm \cite{bocquet_data_2016}.
Analogously to what is done for $\boldsymbol{\Gamma} _k$, each column $i$ of the anomaly matrix $\boldsymbol{\Lambda} _k \in \mathbb{R}^{n_o \times n_e}$ is computed. This matrix also takes into account the discrepancy between the model results and their ensemble average. However, it is defined on the solution space of the observation, and it relies on the projection operator $\mathcal{H}$:

\begin{align}
    \boldsymbol{\Lambda}_{i,k}^f = \frac{\mathcal{H}( \boldsymbol{x}_{i,k}^f ) - \overline{ \mathcal{H}( \boldsymbol{x}_{k}^f ) }}{\sqrt{n_e - 1}} \, , \quad \overline{ \mathcal{H}( \boldsymbol{x}_k^f) } = \frac{\sum_{i=1}^{n_e} { \mathcal{H}( \boldsymbol{x}_{i,k}^f) }}{n_e}
\end{align}

The Kalman gain matrix $\boldsymbol{K}_k$, which describes and measures the correlations between the observations and the state vector, is obtained as follows:

\begin{align}
    \boldsymbol{K}_k = \bm{\Gamma}_k^f \left( \boldsymbol{\Lambda}_k^f \right)^{\top} \left[ \boldsymbol{\Lambda}_k^f \left( \boldsymbol{\Lambda}_k \right)^{\top} + \boldsymbol{\varsigma}_k \right]^{-1}
    \label{eq:KalmanGain}
\end{align}

Finally, the physical state for each ensemble member $i$ is obtained updating the forecast results with a correction term controlled by the Kalman gain:

\begin{equation}
    \boldsymbol{x}_{i,k}^a = \boldsymbol{x}_{i,k}^f + \boldsymbol{K}_k \left( \boldsymbol{y}_{i,k} - \mathcal{H}(\boldsymbol{x}_{i,k}^f) \right) \label{eq:Aassimilation}
\end{equation}

The Kalman filter can be used to augment the state prediction obtained via the model as well as to optimize its free coefficients. In this way, the discrepancy between the model prediction and the observations is naturally reduced. Several strategies are presented in the literature to this purpose, and the model chosen for the present analysis is the \textit{extended state} \cite{Asch2016_SIAM}. In this model, the free parameters of the model are organized in an array $\boldsymbol{\theta}$, which is combined with the state $\boldsymbol{x}$ to obtain an extended state $\boldsymbol{x}^\star$:
\begin{equation}
\boldsymbol{x}^\star = \begin{bmatrix} \boldsymbol{x} \\ \boldsymbol{\theta} \end{bmatrix}    
\end{equation}

The EnKF is then resolved for the extended state $\boldsymbol{x}^\star$, so that the parameters of the model are updated with the forecast of the solution. The steps of the stochastic EnKF used in the present analysis are reported in \autoref{alg:EnKF}.
\begin{algorithm}
    \caption{Algorithm for the stochastic Ensemble Kalman Filter}
    \label{alg:EnKF}
    \textbf{Input:} $\mathcal{M}$, $\mathcal{H}$, $\boldsymbol{\varsigma}_{k+1}$, and some priors for the state system $\boldsymbol{x}_{i,0}^a$, where $\boldsymbol{x}_{i,0}^a \sim \mathcal{N}\left(\boldsymbol{\mu}_x, \boldsymbol{\varsigma}_x^2\right)$. \\
    \For{$k = 0$ to $K-1$}{
        \For{$i = 1$ to $n_e$}{
    \nl Advancement in time of the state vectors:\\
    \qquad $\boldsymbol{x}_{i,k}^f = \mathcal{M}(\boldsymbol{x}_{i,k})$ \\
    \nl Creation of an observation matrix perturbing the observation with a Gaussian noise:\\
    \qquad $\boldsymbol{y}_{i,k} = \boldsymbol{y}_{k} + \bm{\varepsilon}_{i}$, with $\bm{\varepsilon}_{i} \thicksim \mathcal{N}(0,\boldsymbol{\varsigma}_{k})$ \label{line:EnKF_observations} \\
    \nl Projection of the model solution in the observation space:\\
    \qquad $\mathcal{H}(\boldsymbol{x}_{i,k}^f)$\\
    \nl Calculation of the ensemble means:\\
    \qquad $\overline{\boldsymbol{x}}_{k}^f = \frac{1}{n_e} \sum_{i = 1}^{n_e} \boldsymbol{x}_{i,k}^f$,\,
    $\overline{\mathcal{H}(\boldsymbol{x}_{k})} = \frac{1}{n_e} \sum_{i = 1}^{n_e} \mathcal{H}(\boldsymbol{x}_{i,k})$ \\
    \nl Calculation of the anomaly matrices:\\
    \qquad $\boldsymbol{\Gamma}_{k} = \frac{\boldsymbol{x}_{i,k} - \overline{\boldsymbol{x}}_{k}}{\sqrt{n_e-1}}$,\,
    $\boldsymbol{\Lambda}_{k} = \frac{\mathcal{H}(\boldsymbol{x}_{i,k}^f) - \overline{\mathcal{H}(\boldsymbol{x}_{k}^f)}}{\sqrt{n_e-1}}$ \\
    \nl Calculation of the Kalman gain: \\
    \qquad $\boldsymbol{K}_{k} = \boldsymbol{\Gamma}_{k}^f(\boldsymbol{\Lambda}_{k})^{\top} \left[\boldsymbol{\Lambda}_{k}(\boldsymbol{\Lambda}_{k})^{\top} + \boldsymbol{\varsigma}_{k}\right]^{-1}$ \\
    \nl Update of the state matrix:\\
    \qquad $\boldsymbol{x}_{i,k}^a = \boldsymbol{x}_{i,k}^f + \boldsymbol{K}_{k}(\boldsymbol{y}_{i,k}- \mathcal{H}(\boldsymbol{x}_{i,k}))$\\
    } 
    }
\end{algorithm}

\section{Assessment of the DDES model and sensitivity analysis} \label{sec:assesment}

In this chapter, the DDES model used for numerical simulation is validated. In particular, the accuracy of the results will be tested assessing the sensitivity of the model to variations in the computational grid as well as to inlet conditions.

\subsection{Sensitivity to grid refinement}

\begin{figure}[h!]
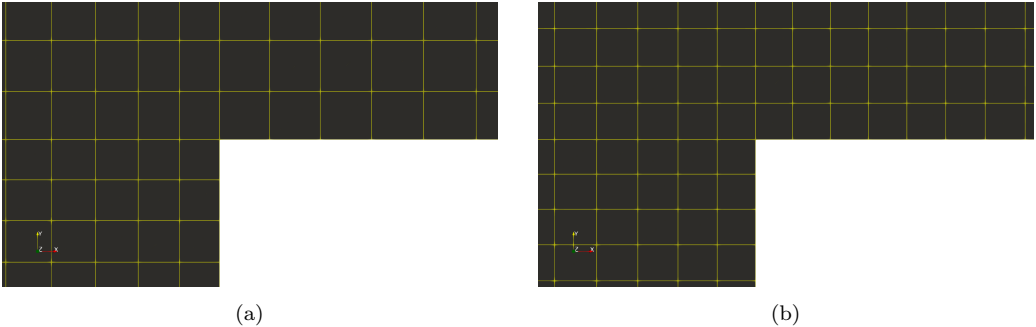

\minipage{0.48\textwidth}
  \includesvg[width=\linewidth]{images/mesh_4M_v3.svg}
  \subcaption{}\label{fig:mesh_4M}
\endminipage\hfill
\minipage{0.48\textwidth}%
  \includesvg[width=\linewidth]{images/mesh_30M_v4.svg}
  \subcaption{}\label{fig:mesh_30M}
\endminipage
\caption{Zoom of the mesh resolution close to the leading edge for grids (a) G1 and (b) G2. $x \in [-2.54, -2.44]$ and $y \in [0.47, 0.525]$.} \label{fig:mesh_resolutions}
\end{figure}

Two DDES runs, referred to as DDES-I0-G1 and DDES-I0-G2, are performed using a constant velocity inlet and grids of different resolution. The grids, whose details are reported in \autoref{tab:Grids}, are referred to as G1 and G2, respectively, and their resolution in the proximity of the leading edge is shown in \autoref{fig:mesh_resolutions}. The main difference between the two grids is their global resolution, which sums up to 4 million elements and 30 million elements. On the other hand, the strategy for the distribution of the mesh elements is very similar for G1 and G2, and it follows indications of previous CFD works in the literature \cite{mariotti_stochastic_2017,cimarelli_direct_2018,rocchio_flow_2020}. Owing to the statistical homogeneity of the flow in the spanwise direction $z$, the size of the mesh elements is constant and equal to $\Delta_z^\star=\Delta_z / D$. In the streamwise direction $x$ and the normal direction $y$, the smallest grid elements are in correspondence of the two leading edges and the two trailing edges. In these locations, the size of the mesh elements is $\Delta_x^\star$ and $\Delta_y^\star$, respectively. Moving away from the edges, the size of the mesh elements increases following a geometric progression of ratio $r_x$ and $r_y$. In the wake region, for $x=5$, $y=0$ the resolution of the mesh is equal to $\Delta_x^{d}$ and $\Delta_y^{d}$.

\begin{table}[h!]
\centering
\begin{tabular}{lcccccccc} \toprule[1.5pt]
CFD run & $\Delta_x^\star$ & $\Delta_y^\star$ & $\Delta_z^\star$ & $r_x$ & $r_y$ & $\Delta_x^{d}$ & $\Delta_y^{d}$ & $N$ \\
G1 & 0.01 & 0.012 & 0.1 & 1.012 & 1.059 & 0.063 & 0.016 & $4 \cdot 10^6$  \\
G2 & 0.0074 & 0.009 & 0.05 & 1.012 & 1.015 & 0.031 & 0.014 & $30 \cdot 10^6$ \\
\bottomrule[1.5pt]
\end{tabular}
\caption{Characteristics of the grids used for the numerical simulations. The size of the elements is provided in $D$ units.}
\label{tab:Grids}
\end{table}

\begin{figure}
    \centering
    \begin{tabular}{cc}
        \includesvg[width=.49\textwidth]{images/UMean_4M_SA-DDES_vMe2.svg} & \includesvg[width=.49\textwidth]{images/UMean_30M_SA-DDES_vMe2.svg} \\
         (a) & (b) \\
         \includesvg[width=.49\textwidth]{images/UMean_rans2D_vMe2.svg} &  \includegraphics[width=.49\textwidth]{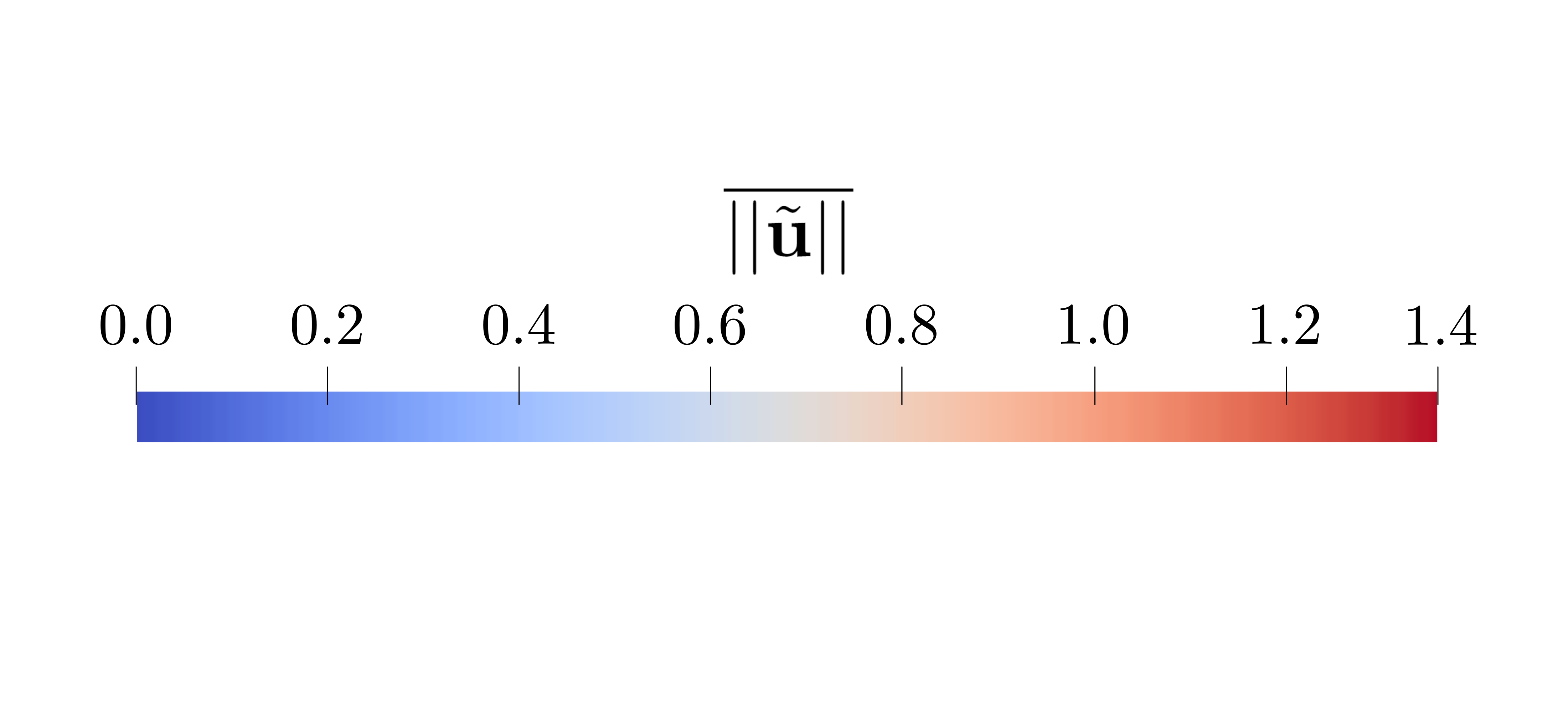} \\
         (c) & 
    \end{tabular}
    \caption{Isocontours of the time-averaged velocity magnitude $\overline{||\mathbf{\tilde{u}}||}$. Results are shown for the simulations (a) DDES-I0-G1, (b) DDES-I0-G2 and (c) a RANS calculation.}
    \label{fig:UMean_DDES}
\end{figure}

The results from the simulations DDES-I0-G1 and DDES-I0-G2 are compared with those obtained by a RANS run using the Spalart-Allmaras model. The grid used for this last calculation is identical to G1 in the $x$ and $y$ directions but it is two-dimensional. Isocontours of the time-averaged velocity magnitude are shown in \autoref{fig:UMean_DDES}. The three simulations are capturing the main physical aspects of the flow which include separation at the leading edge, the formation of a recirculation bubble, flow reattachment and evolution of a wake region downstream. However, results from the RANS calculation indicate a shorter recirculation bubble. The reattachment of the flow happens here at $65\%$ of the total length L, against $75\%$ for DDES-I0-G1 and $76\%$ for DDES-I0-G2. The results for the two DDES runs is qualitatively in line with experimental observations from CSTB for the same geometry, Reynolds numbers and low turbulence-intensity upstream conditions. Depending on the value of the turbulence intensity upstream, experiments show a flow reattachment in the range $50\%$ to $80\%$ approximately, with lower turbulence intensity providing larger recirculation bubbles \cite{Bruno2014BenchmarkOT}. The results obtained with the DDES here performed are towards the larger values of the experimental range. Considering that the inlet condition here used does not include any synthetic turbulence model, results obtained for this case, as it will shown further, are encouraging and constitute a good starting field to initialize the solution during the sequential Data Assimilation approach.

\begin{figure}[h!]
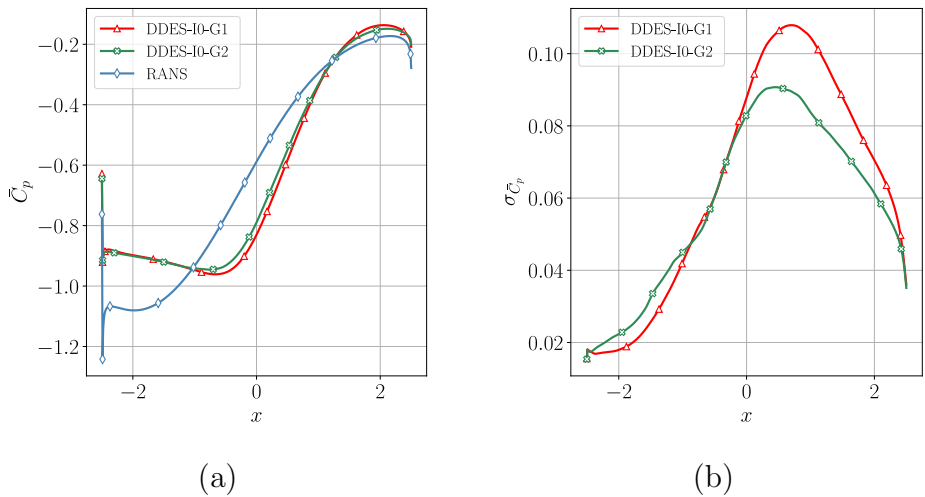

\centering
    \begin{tabular}{cc}
        \includesvg[width=.45\textwidth]{images/Cpm_4M_30M_v6.svg} & \includesvg[width=.45\textwidth]{images/Cpstd_4M_30M_v6.svg} \\
        (a) & (b)
    \end{tabular}
  \caption{Pressure coefficient $C_p$ at the body surface. (a) The time-averaged distribution $\overline{C}_p$ and (b) the variance $\sigma_{\bar{C}_p}$.}
  \label{fig:Cp_DDES}
\end{figure}

The pressure coefficient $C_p = 2(p - p_\infty) / \rho U_\infty^2$ is now analyzed. The time average and the variance of $C_p$ at the body surface is shown for the three simulations in \autoref{fig:Cp_DDES}. The variance for the RANS calculation is not available, as a steady-state simulation was performed. For incompressible flow simulation, results from the analysis of the features of the pressure field should be carefully interpreted. In fact, the variable $p$ is a Lagrangian multiplier that is manipulated to obtain a solenoidal condition for the velocity field. Nonetheless, it can usually provide a reliable map of the flow condition at the wall. One can see again that a significant discrepancy is observed between the RANS and the DDES simulations for the time-averaged $C_p$. On the other hand, runs DDES-I0-G1 and DDES-I0-G2 exhibit very similar results. The peak of the variance of $C_p$ is observed for a streamwise position of $\approx 65\% \, L$ i.e. around $10\%$ before than the flow re-attachment. This result is consistent with experiments and numerical simulations in the literature \cite{Bruno2014BenchmarkOT,rocchio_flow_2020}.

\begin{figure}[h!]
    \includesvg[width=.75\textwidth]{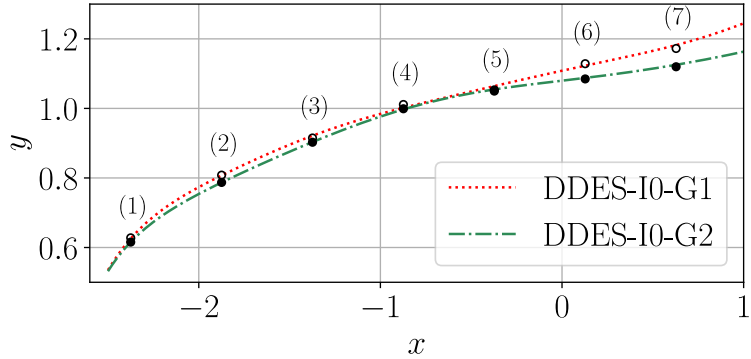}
    \caption{Position of the sensors used to perform the spectral analysis of the velocity field obtained via DDES. Empty circles represent DDES-I0-G1 probes positioning, full circles represent DDES-I0-G2 probes positioning.}
  \label{fig:LocSensorMorlet_DDES}
\end{figure}

\begin{figure}[h!]
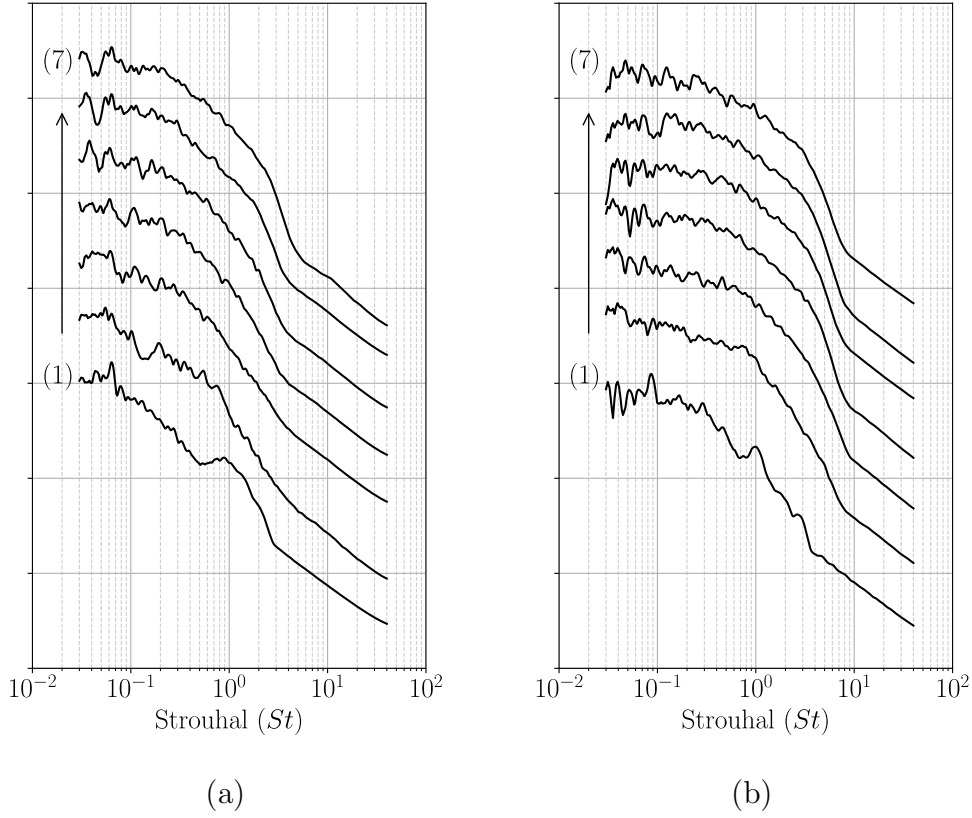

    \begin{tabular}{cc}
        \includesvg[width=0.48\textwidth]{images/morlet_4M_vfinale_december_2.svg} &  \includesvg[width=0.48\textwidth]{images/morlet_30M_vfinale_december_2.svg} \\
         (a) & (b) 
    \end{tabular}
  \caption{Spectra obtained via Morlet transform for simulations (a) DDES-I0-G1 and (b) DDES-I0-G2. Results from different sensors are shown using an offset to improve the clarity of the representation.}
  \label{fig:SpectraMorlet_DDES}
\end{figure}

At last, features of the instantaneous flow are analyzed. Time resolved velocity for the runs DDES-I0-G1 and DDES-I0-G2 is sampled in correspondence of 7 sensors. These sensors are located on the outer limit of the detaching shear layer i.e. where the mean velocity exhibits its maximum (see the procedure used in \citet{rocchio_flow_2020, moldovan2022multigrid}). The sensors locations are listed in \autoref{tab:probes_location}. \autoref{fig:LocSensorMorlet_DDES} shows the detaching layers for runs DDES-I0-G1 and DDES-I0-G2. This sampled field is used to obtain spectra via a Morlet transform. The spectra, which are shown in \autoref{fig:SpectraMorlet_DDES}, again exhibit minimal differences. One can see that a peak for the spectrum obtained for the first probe with $x < -2$ is clearly observed for $St \approx 1$, where $St$ is the Strouhal number. This peak, which is associated with the frequency of Kelvin--Helmholtz instabilities, has also been observed in LES runs \cite{Rocchio2020}.  

\begin{table}[h!]
\centering
\begin{tabular}{l c l c } \toprule[1.5pt]
DDES-I0-G1 & & DDES-I0-G2 & \\ \midrule[0.75pt]
probe (1) & (-2.374, 0.628) & probe (1) & (-2.374, 0.616) \\
probe (2) & (-1.874, 0.808) & probe (2) & (-1.874, 0.787) \\
probe (3) & (-1.374, 0.915) & probe (3) & (-1.374, 0.903) \\
probe (4) & (-0.873, 1.011) & probe (4) & (-0.873, 0.999) \\
probe (5) & (-0.373, 1.05)  & probe (5) & (-0.373, 1.053) \\
probe (6) & (0.128, 1.129)  & probe (6) & (0.128, 1.085) \\
probe (7) & (0.628, 1.172)  & probe (7) & (0.628, 1.120) \\
\bottomrule[1.5pt]
\end{tabular}
\caption{Position of the sensors used to perform the Morlet transform of the velocity field obtained via DDES.}
\label{tab:probes_location}
\end{table}

In summary, the DDES simulations provide a very similar prediction, which appears to be significantly improved when compared with the RANS calculation. For all of these reasons, the grid G1 is chosen to perform the simulations for the present work.

\subsection{Sensitivity to the inlet conditions}
\label{sec::sensitivityInlet}

A third numerical simulation, referred to as DDES-I1-G1, is run using the synthetic turbulence inlet presented in \autoref{sec3:setup}. The values of the eight coefficients are mainly selected according to the experiments performed at CSTB. More precisely, the components of the Reynolds stress tensor $\tau_{t,ij}$ and the integral length scale $\ell$ are set according to measurements performed in the free wind tunnel i.e. before the rectangular cylinder was installed. The parameter $\delta$, which is related to the mesh size in the proximity of the boundary condition, is set to an average value of $0.115$. The values for the eight parameters are summarized in line 1 of table \autoref{tab:cstb_param}.

\begin{table}[h!]
\centering
\resizebox{\textwidth}{!}{
\begin{tabular}{lcccccccc} \toprule[1.5pt]
 & $\tau_{t,xx}$ & $\tau_{t,yy}$ & $\tau_{t,zz}$ & $\tau_{t,xy}$ & $\tau_{t,xz}$ & $\tau_{t,yz}$ & $\ell$ & $\delta$ \\
Exp. reference & $1.25 \cdot 10^{-3}$   & $0.99\cdot 10^{-3}$ & $0.93\cdot 10^{-3}$ & $-2.00\cdot 10^{-5}$ & $5.00\cdot 10^{-5}$ & $-3.00\cdot 10^{-5}$ & 1.38 & -- \\
DA - prior     & $5.65 \cdot 10^{-5}$  & $5.10\cdot 10^{-5}$   & $4.05\cdot 10^{-5}$ & $-9.55\cdot 10^{-7}$ & $2.45\cdot 10^{-6}$ & $-1.41\cdot 10^{-6}$ & $1.04$ &  $0.115$\\
DA - optimized & $5.34 \cdot 10^{-5}$   & $4.18 \cdot 10^{-5}$ & $7.13 \cdot 10^{-5}$ & $-7.04 \cdot 10^{-7}$ & $2.99 \cdot 10 ^{-6}$ & $-1.04 \cdot 10^{-6}$ & $1.199$ & $0.0606$  \\
\bottomrule[1.5pt]
\end{tabular}}
\caption{Parametric description of the synthetic turbulence inlet condition used in this work.}
\label{tab:cstb_param}
\end{table}

\begin{figure}
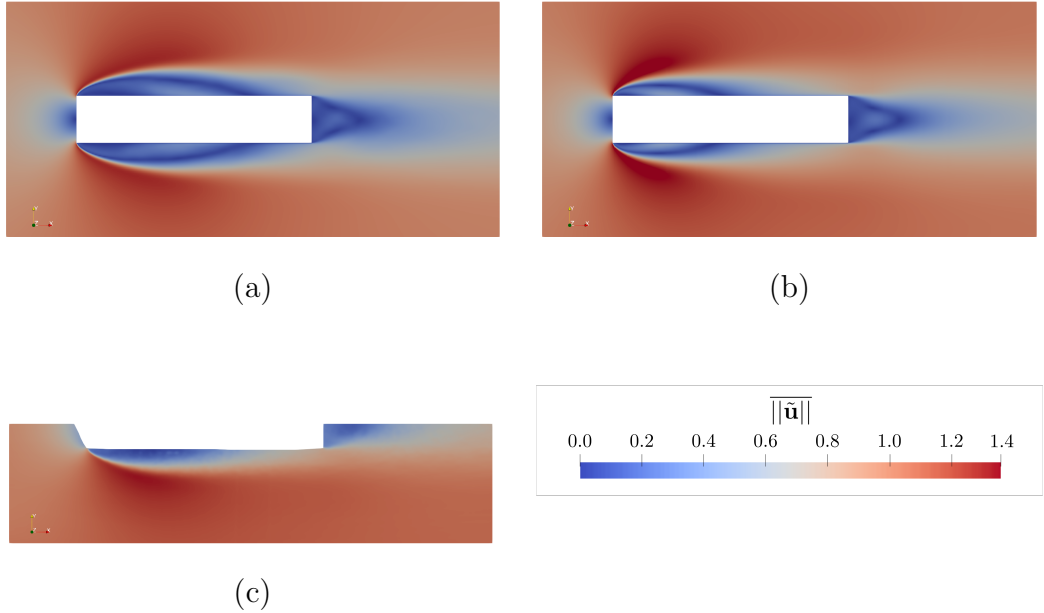

    \centering
    \begin{tabular}{cc}
        \includesvg[width=.49\textwidth]{images/UMean_4M_SA-DDES_vMe2.svg} & \includesvg[width=.49\textwidth]{images/UMean_paramCSTB_vMe2.svg} \\
         (a) & (b) \\
         \includesvg[width=.49\textwidth]{images/UMean_CSTB_v6.svg} &  \includegraphics[width=.49\textwidth]{images/0_UMean_SCALE_big_ok_ok.png} \\
         (c) & 
    \end{tabular}

    \caption{Isocontours of the time-averaged velocity magnitude $\overline{||\mathbf{\tilde{u}}||}$. Results are shown for the simulations (a) DDES-I0-G1, (b) DDES-I1-G1 and (c) CSTB experiments.}
    \label{fig:UMean_DDES_CSTB_exp_ini}
\end{figure}

\begin{figure}[h!]
    \centering
    \includesvg[width=.8\textwidth]{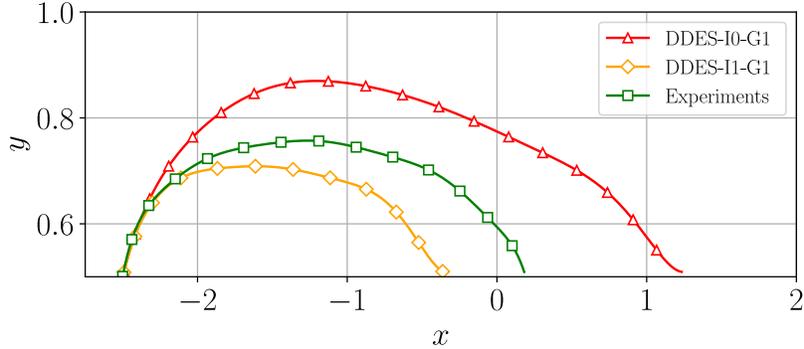}
    \caption{Recirculation region at the top of the rectangular cylinder.}
    \label{fig:RecRegionBeforeDA}
\end{figure}

Results from the numerical simulations DDES-I0-G1 and DDES-I1-G1 are compared with the experimental results obtained at the wind tunnel of CSTB. Isocontours of the velocity magnitude are shown in \autoref{fig:UMean_DDES_CSTB_exp_ini} and the size of the recirculation region is shown in \autoref{fig:RecRegionBeforeDA}. One can see significant discrepancies between numerical and experimental results and in particular both numerical simulations do not provide an accurate representation of the recirculation bubble. This result was expected for the simulation with the classical, zero turbulence intensity inlet. However, one can see that the usage of the numerical values in \autoref{tab:cstb_param} with the synthetic turbulent inlet implemented in OpenFOAM produces a higher level of turbulence intensity than expected. The reasons for this results can be associated to the structural features of the inlet model and its interactions with the mesh resolution and turbulence model. A minor effect could also be associated with uncertainties in the experimental measurements. One can also see that both numerical simulations develop a secondary bubble at the leading edge. This result, which is not observed in the experiments, is actually pretty common in scale resolved numerical simulation for this test case \cite{Bruno2014BenchmarkOT}.   

\begin{figure}[h!]
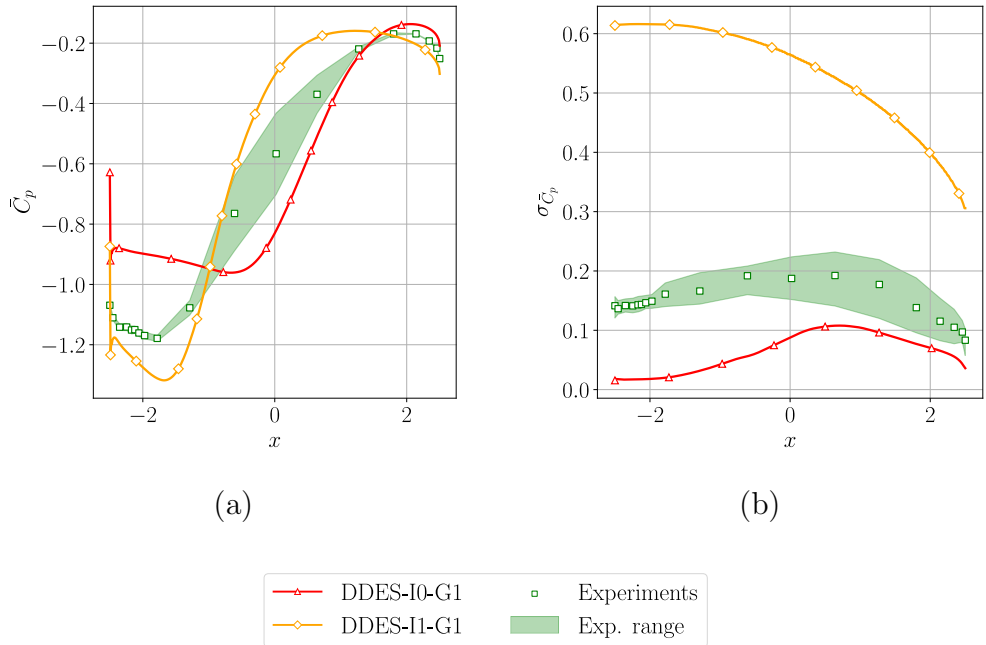

    \begin{tabular}{cc}
        \includesvg[width=0.48\textwidth]{images/Cpm_final_new3_sizetest.svg} & \includesvg[width=0.48\textwidth]{images/Cpstd_final_nolim_ok_1_sizetest.svg} \\
        (a) & (b) \\
        &  \\
    \end{tabular}
    \includesvg[height=0.085\textwidth]{images/Cpstd_legend_1.svg}
    \caption{Features of the pressure field at the top wall. Distributions of (a) the time-averaged pressure coefficient $\overline{C}_p$ and (b) its variance are shown. Numerical results obtained from the DDES simulations are compared with experimental data.}
    \label{fig:Cpm_Cpstd_p2}
\end{figure}

Similar conclusions can be drawn via the analysis of the time-averaged $C_p$ and its variance, which are shown in \autoref{fig:Cpm_Cpstd_p2}. Significant discrepancies are observed for the distribution of the mean pressure coefficient. In addition, the magnitude of the variance of the pressure for the simulation DDES-I1-G1 is approximately four times larger than the experimental results. This observation is partially due to extreme peaks of the pressure that are locally observed. These peaks are most likely induced by the interaction of the numerical solver, the zero-gradient boundary condition at the wall and the unsteady behavior produced by the inlet. However, results in \autoref{fig:Cpm_Cpstd_p2} (b) are filtered to exclude the larger peaks numerically observed. Therefore, one could argue that the distribution shown is mainly associated with the inlet boundary condition used for simulation DDES-I1-G1. The red line indicating the variance of the simulation DDES-I0-G1 is significantly lower than the experimental reference, in particular close to the leading edge. This result indicates that the classical inlet used for this simulation fails to capture important phenomena at play close to the separation due to the sharp edges.  

In summary, present results indicate that the usage of a synthetic turbulence inlet condition significantly affect the flow organization, and it can have the potential to produce accurate results in comparison with the experiments. However, the parametric set-up of such inlet can not be directly extrapolated from the data sampled in the wind tunnels, because of strong non-linear interactions between several sources of error present in the CFD solver. In the following, a data-driven procedure will be used to calibrate the value of the model constants, in order to minimize the discrepancy observed with experiments.

\section{Data-driven augmentation of inlet boundary conditions}
\label{sec:DA-results}

Discussion in \autoref{sec::sensitivityInlet} highlighted the numerical difficulties in obtaining an accurate inlet condition and how this affects the comparison with experimental data. In this section, an EnKF approach is used to optimize the parametric description of the synthetic turbulence inlet used in OpenFOAM in order to reduce the discrepancy between numerical and experimental results. 

\subsection{Selection of the EnKF hyperparameters and DA experiment}

The EnKF relies on two main sources of information:
\begin{itemize}
\item A \textit{model}, which provides a quasi-continuous description of the flow configuration investigated. The model chosen for the present investigation is the DDES CFD solver, which calculates the instantaneous evolution of the velocity and pressure fields.
\item Some \textit{observation}, which is usually high-fidelity data sparse in space and time. In this case, experimental results from CSTB are used. These results are obtained from $451$ sensors which are positioned on the PIV plane normal to the spanwise direction $z$. The positioning of the sensors is shown in \autoref{fig:SensorsExp}. On this location, the time-averaged values of the streamwise velocity $u_x$ and the normal velocity $u_y$ are sampled. 
\end{itemize}

One can see that the CFD model and the experimental observation are different in nature i.e. the model prediction provides instantaneous information while experimental results are time-averaged. Therefore, the DA strategy must be adapted to take into account this difference. The algorithm is organized in five different steps:
  
\begin{description}
    \item[$(o)$] The $n_e=30$ DDES calculations which constitute the numerical ensemble used in the EnKF are initialized with random values for the parametric description of the inlet. These parameters include the six component $\tau_{t,ij}$ of the Reynolds stress tensor at the inlet, as well as the integral length scale $\ell$ and the mesh characteristic length $\delta$. These last two parameters are used to determine the length scale $\sigma _i$ in \autoref{eq:alpha_sigma}. Using the formalism of Data Assimilation, the free coefficients are arranged in an array $\boldsymbol{\theta}$:
    \begin{align}
    \boldsymbol{\theta}_t =\begin{bmatrix}
           \tau_{t,xx} \\
           \tau_{t,yy} \\
           \tau_{t,zz} \\
           \tau_{t,xy} \\
           \tau_{t,xz} \\
           \tau_{t,yz} \\  
           \end{bmatrix} \qquad
    \boldsymbol{\theta}_l =\begin{bmatrix}
           \ell \\
           \delta 
         \end{bmatrix} \qquad
    \boldsymbol{\theta} = \begin{bmatrix}
    \boldsymbol{\theta}_t \\
    \boldsymbol{\theta}_l \\
    \end{bmatrix}
  \end{align}

The random values for each of the coefficients of $\bm{\theta}$ and for each of the simulations is determined via sampling of a truncated Gaussian probability density function $\mathcal{N}(\mu_{\theta_i},\varsigma_{\theta_i}^2)$. $\mu$ and $\varsigma$ are the average and the standard deviation provided for each parameter $\theta_i$. These parameters, which are reported in \autoref{tab:state}, have been chosen taking into account the experimental results provided by CSTB and after a careful investigation of the preliminary CFD results. The choice of the prior values for the parametric description is important to obtain a fast and robust convergence of the EnKF optimization \cite{villanueva_augmented_2023}. Values for the parameters are accepted within the range $[\mu - 2 \varsigma, \, \mu + 2 \varsigma]$ and are re-sampled if they lay outside the prescribed interval.
\begin{table}[h!]
\centering
\resizebox{\textwidth}{!}{
\begin{tabular}{ccccccccc} \toprule[1.5pt]
 & $\tau_{t,xx}$ & $\tau_{t,yy}$ & $\tau_{t,zz}$ & $\tau_{t,xy}$ & $\tau_{t,xz}$ & $\tau_{t,yz}$ & $\delta$ & $\ell$ \\
 $\boldsymbol{\mu}_x$ & $5.652\cdot 10^{-5}$ & $5.105 \cdot 10^{-5}$ & $4.046 \cdot 10^{-5}$ & $-9.553\cdot10^{-7}$ & $2.446\cdot 10^{-6}$ &  $-1.411 \cdot 10^{-7}$ &  $9.22 \cdot 10^{-2}$ & $1$ \\
 $\boldsymbol{\varsigma}_x^2$ & $1.72\cdot 10^{-10}$ & $9.02 \cdot 10^{-11}$ & $6.4\cdot10^{-11}$ & $4.13\cdot10^{-14}$ & $2.07\cdot10^{-13}$ & $9.34\cdot10^{-14}$ & $1.39\dot10^{-3}$ & $3.98\cdot10^{-2}$ \\
\bottomrule[1.5pt]
\end{tabular}}
\caption{Features of the probabilistic distribution for the prior parameters of the CFD ensemble members.}
\label{tab:state}
\end{table}
The instantaneous physical state for $t=0$ is the same for all the ensemble members and it is obtained by a preliminary DDES run with the mean parameters $\mu$ in \autoref{tab:state}. 

    \item[$(i)$] The $n_e$ DDES simulations are run for a total of $t=25\,t_A$ times, where the advection time scale is $t_A = D / U_\infty$. This time window is divided into two phases. For $t < 10\,t_A$, no average is performed as the effects of the inlet parametric description are advected in the domain. The threshold $10\,t_A$ has been chosen observing the results of prior runs, for which aerodynamic coefficients would stabilize after $5\,t_A$ to $7\,t_A$. In the second phase for $t \in [10,25]\,t_A$, averages are performed in time and in the spanwise direction. This interval has been chosen observing the behavior of a preliminary run, where over this time window the rate of convergence of the mean velocity and pressure would be $\approx 5\%$ i.e. the same order of magnitude of the uncertainty in experimental results. This phase corresponds to the \textit{forecast} of the EnKF.
    \item[$(ii)$] After the forecast is performed, the mean velocity field for each ensemble member $i$ is post-processed to obtain $\mathcal{H}(\boldsymbol{x}_{i} ^f)$ i.e. the velocity field is interpolated over the sensors where experimental observation is available. Because of the homogeneity in the spanwise direction, only the components in the streamwise direction $\overline{\tilde{u}_x}$ in the normal direction $\overline{\tilde{u}_y}$ are interpolated. Therefore, the vector including the model information used for DA for the ensemble member $i$ is:
    \begin{align}
        \mathcal{H}(\boldsymbol{x}_i)=
        \begin{pmatrix}
            \overline{\tilde{u}_{x_1}} &
            \dots &
            \overline{\tilde{u}_{x_n}} &
            \overline{\tilde{u}_{y_1}}&
            \dots &
            \overline{\tilde{u}_{y_n}} &
        \end{pmatrix} \in \mathbb{R} ^ { 2 n_o } \label{eq:DA_step2}
    \end{align}
    Where the number of sensors is $n_o = 451$. The complete matrix used for ensemble approximation is $\mathcal{H}(\boldsymbol{x}) \in \mathbb{R}^{902 \times 30}$. The position of the sensors over the average velocity magnitude of one of the ensemble members is shown in \autoref{fig:SensorsExp}.
        
    \begin{figure}[h!]
        \centering
        \begin{tabular}{c}
         \includesvg[width=0.85\textwidth]{images/ObsSetConSimulation.svg}  \\
         \includegraphics[width=0.65\textwidth]{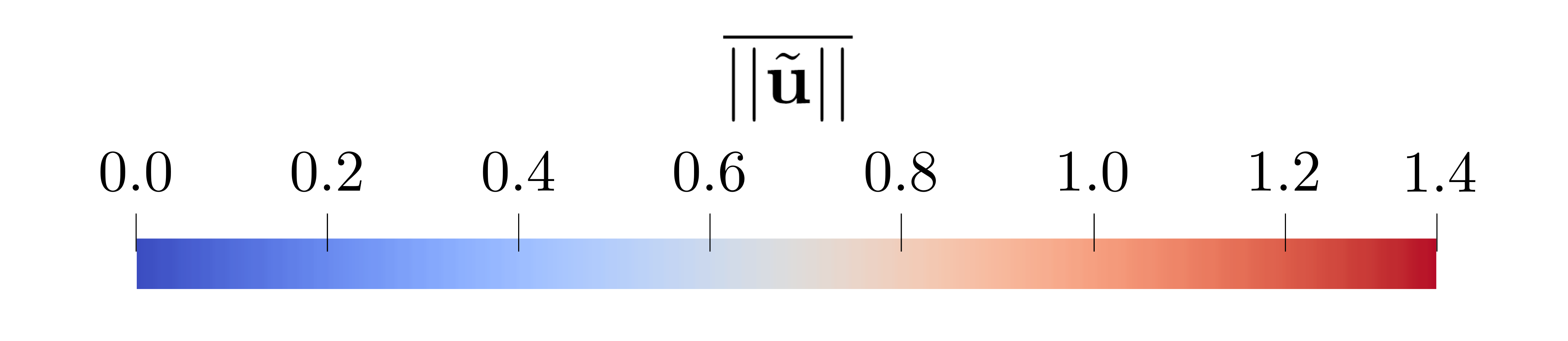}
        \end{tabular}
        \caption{Position of the sensors used to sample the experimental data. The sensors are shown with the velocity magnitude obtained by one of the DDES ensemble members.}
        \label{fig:SensorsExp}
    \end{figure}

    \item[$(iii)$] The other numerical ingredients used in the EnKF are assembled. The observation vector, which is composed by the time-averaged velocity field sampled over the sensors from the experiments, is perturbed adding a Gaussian noise $\mathcal{N}(0, \varsigma_y)$, where $\varsigma_y$ accounts for the estimated $5\%$ uncertainty over the experimental data. Using the classical hypothesis that observation from different sensors is not correlated \cite{bocquet_data_2016}, the additive noise is used to obtain $n_e$ observation vectors which are combined in the observation matrix $\boldsymbol{Y}^{902 \times 30}$. The physical state obtained via model realizations is also used to calculate the anomaly matrices according to \autoref{eq:anomaly_gamma}. All these numerical elements are combined to obtain the Kalman gain $\boldsymbol{K}$ using \autoref{eq:KalmanGain}.
    \item{$(iv)$} The \textit{analysis} phase takes place, where the augmented state $\boldsymbol{x}_i^a$ for each ensemble member $i$ is obtained with \autoref{eq:Aassimilation} from its forecast $\boldsymbol{x}_i^f$, the Kalman gain $\boldsymbol{K}$ and the discrepancy between experiments and model prediction. The latter is measured via the difference of the $i^{\text{th}}$ column of $\boldsymbol{Y}$ and $\mathcal{H}(\boldsymbol{x})$. The physical state correction obtained via DA, which is derived for a steady configuration, is discarded. On the other hand, the EnKF update to the parameters is conserved. However, in order to avoid spurious oscillations of the coefficients which can be produced by the Kalman update for the analysis of strongly unsteady configurations \cite{villanueva_LES_2023}, the maximum / minimum value accepted for the following forecast are truncated to the extremes of the previous parametric range. In order to obtain a sufficient variability of the ensemble, the parameters are then inflated using the same variance that was used for the prior state. Once the parameters are updated, the new forecast is ready to be submitted without any change of the initial conditions for bulk velocity and pressure.
\end{description}

\begin{figure}
    \centering
    \includesvg[width=\textwidth]{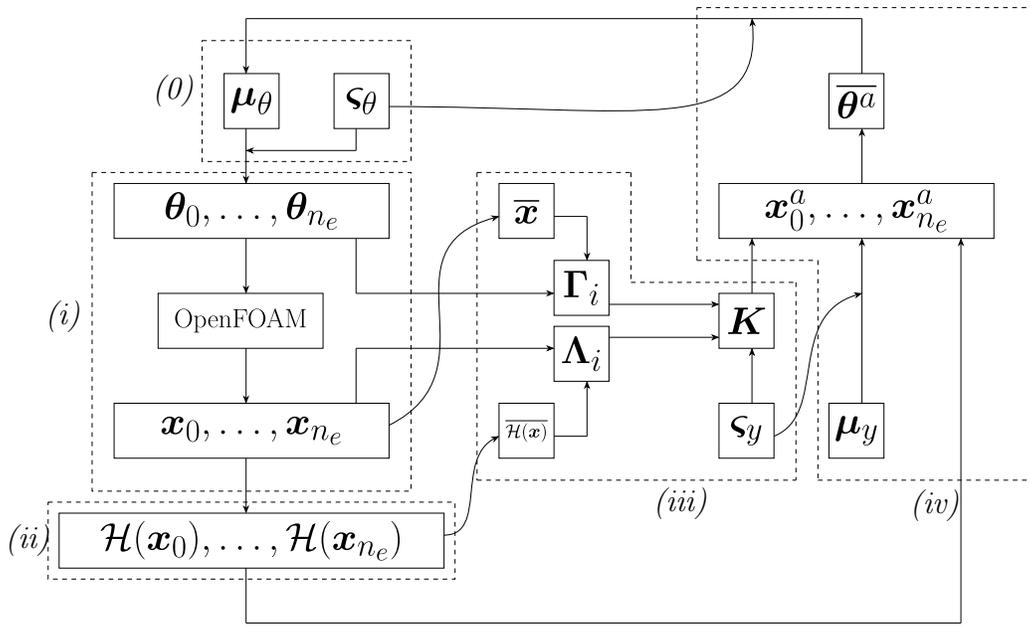}
    \caption{EnKF algorithm scheme.}
    \label{fig:scheme_EnKF_method}
\end{figure}

This algorithm, which is exemplified in the scheme in \autoref{fig:scheme_EnKF_method}, has been here repeated for three complete cycles i.e. three forecasts and three analyses. At the end of the last forecast, the member exhibiting the lowest discrepancy with the experiments, in terms of features of the recirculation bubble, has been selected. A full run has been performed for this member for a total of $300\,t_A$. The results obtained with this optimized DDES run, which will be referred to as DDES-DA-G1 in the following, are going to be compared with simulation DDES-I0-G1, DDES-I1-G1 and the experiments from CSTB.

\subsection{Results}

\begin{figure}
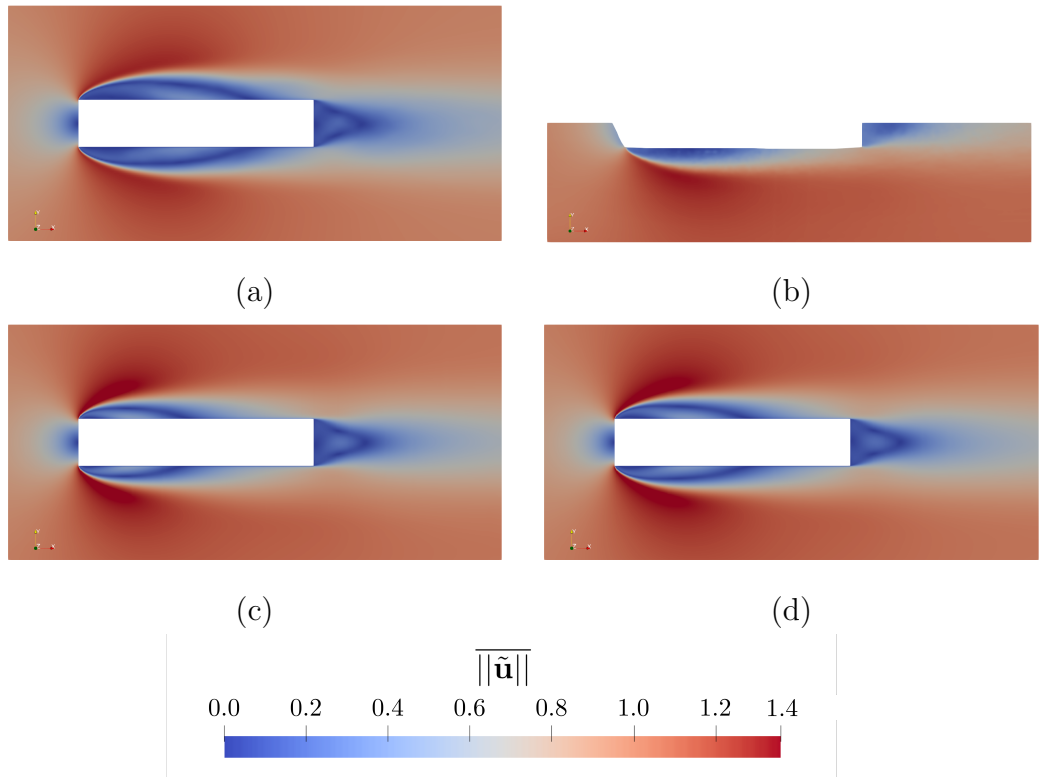

    \centering
    \begin{tabular}{cc}
        \includesvg[width=.49\textwidth]{images/UMean_4M_SA-DDES_vMe2.svg} & \includesvg[width=.49\textwidth]{images/UMean_CSTB_v6.svg} \\
         (a) & (b) \\
         \includesvg[width=.49\textwidth]{images/UMean_paramCSTB_vMe2.svg} &  \includesvg[width=.49\textwidth]{images/UMean_finalParameters_vMe2.svg} \\
         (c) & (d)
    \end{tabular}
    \centering
    \includegraphics[width=0.65\textwidth]{images/0_UMean_SCALE_small_ok.png}
    \caption{Isocontours of the time-averaged velocity magnitude $\overline{||\mathbf{\tilde{u}}||}$. Results are shown for (a) the simulation DDES-I0-G1, (b) experimental data from CSTB's wind tunnel (c) the simulation DDES-I1-G1 and (d) the simulation DDES-DA-G1}
    \label{fig:UMean_DDES_CSTB_exp}
\end{figure}

\begin{figure}[h!]
    \centering
    \includesvg[width=0.8\textwidth]{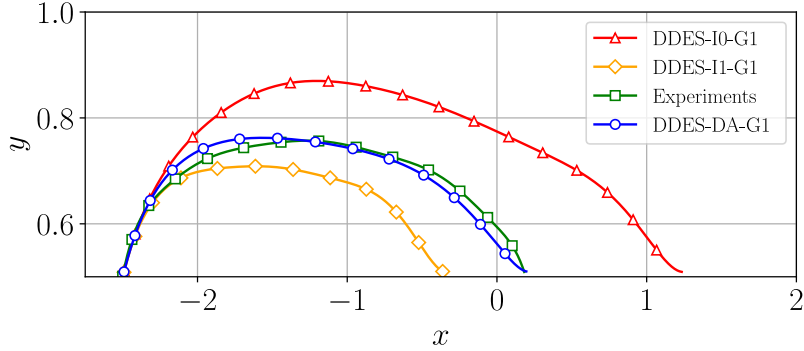}
    \caption{Size of the recirculation bubble for the numerical simulations performed, which are compared with available experimental results.}
    \label{fig:RecBubbleDA}
\end{figure}

The physical prediction of the flow obtained with the DA algorithm is now analysed. First, the isocontours of the velocity magnitude are presented in \autoref{fig:UMean_DDES_CSTB_exp} for a qualitative evaluation. One can see that the data-driven run DDES-DA-G1 convincingly captures the flow topology like the prior simulations DDES-I0-G1 and DDES-I1-G1. However, the flow features in the proximity of the recirculation region appear to be in better agreement with the experimental data. This qualitative observation is further confirmed by the analysis of the size of the recirculation bubble, which is shown in \autoref{fig:RecBubbleDA}. One can see that the curve obtained by the simulation DDES-DA-G1 almost exactly superposes with the experimental data with a significant improvement in the accuracy when compared with the two DDES preliminary simulations.

\begin{figure}[h!]
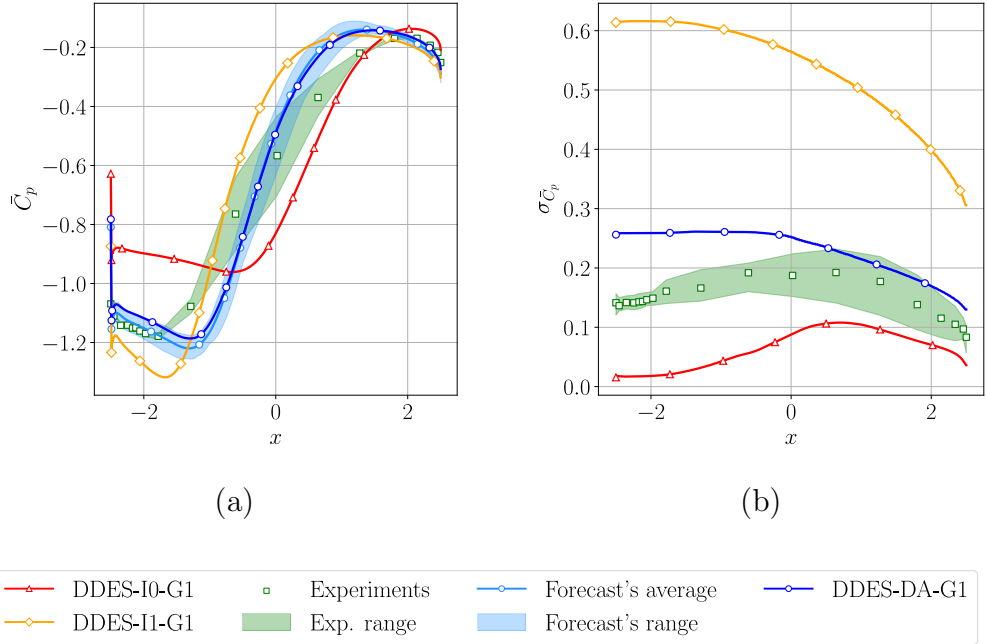

    \centering
    \begin{tabular}{cc}
        \includesvg[width=0.48\textwidth]{images/Cpm_final_new4_sizetest.svg} & \includesvg[width=0.48\textwidth]{images/Cpstd_final_nolim_ok_2_sizetest.svg} \\
        (a) & (b) \\
         & 
    \end{tabular}
    \includesvg[height=0.085\textwidth]{images/Cpm_legend_2.svg}
    \caption{Features of the pressure field at the top wall. Distributions of (a) the time-averaged pressure coefficient $\overline{C}_p$ and (b) its variance are shown. Numerical results obtained from the DDES simulations are compared with experimental data and the optimized DA run.}
    \label{fig:Cpm_Cpstd_p3}
\end{figure}

The features of the pressure field on the top wall of the rectangular cylinder are now analysed. These quantities are of particular interest, because experimental results are available but not used as available information for the DA analysis step. Therefore, this comparison can highlight the global improvement in accuracy of the data-driven method. The time-averaged $\overline{C}_p$ and its variance $\sigma_{\overline{C}_p}$ are shown in \autoref{fig:Cpm_Cpstd_p3}. The blue shaded area corresponds to the variability of the solution of the DA final forecast, with the simulation DDES-DA-G1 in dark blue. On the other hand, the green shaded area corresponds to the confidence in the local experimental measurements. One can see that the prediction of $\overline{C}_p$ for the DA model, shown in  \autoref{fig:Cpm_Cpstd_p3} (a), exhibits a lower discrepancy with experimental data when compared with prior DDES runs. In addition, one can see a good superposition of the experimental and data-driven shaded areas, which indicates that the application of the EnKF was robust. Similarly to the time-averaged $\overline{C}_p$, the variance $\sigma_{\overline{C}_p}$ for the DDES-DA-G1 run shows closer values to the experiments. The DA run, again represented in dark blue, is in very good agreement with the experimental data from approximately the re-attachment point of the recirculation bubble to the trailing edge. In the recirculation bubble region, $\sigma_{\overline{C}_p}$ is significantly higher than the experiments, but still provides the lower normalized discrepancy for the three numerical simulations performed. This discrepancy can be again qualitatively associated with peaks of the pressure field which results from the interaction of boundary conditions and the numerical solvers. Contrarily to simulation DDES-I1-G1, the filtering of the results affects the DA profile for around $2\%$-$3\%$ of the magnitude, indicating that this issue is not the main mechanism at play for the determination of the pressure variance for the run DDES-DA-G1.

\begin{figure}[h!]
    \begin{tabular}{cc}
         \includesvg[width=0.48\textwidth]{images/UPrime2Mean_XY_4M_SA-DDES_v4.svg} & \includesvg[width=0.48\textwidth]{images/UPrime2Mean_XY_CSTB_v6.svg} \\
         (a) & (b) \\
         \includesvg[width=0.48\textwidth]{images/UPrime2Mean_XY_paramCSTB_v4.svg} & \includesvg[width=0.48\textwidth]{images/UPrime2Mean_XY_finalParameters_v4.svg} \\
         (c) & (d) \\
    \end{tabular}
    \includegraphics[width=.65\textwidth]{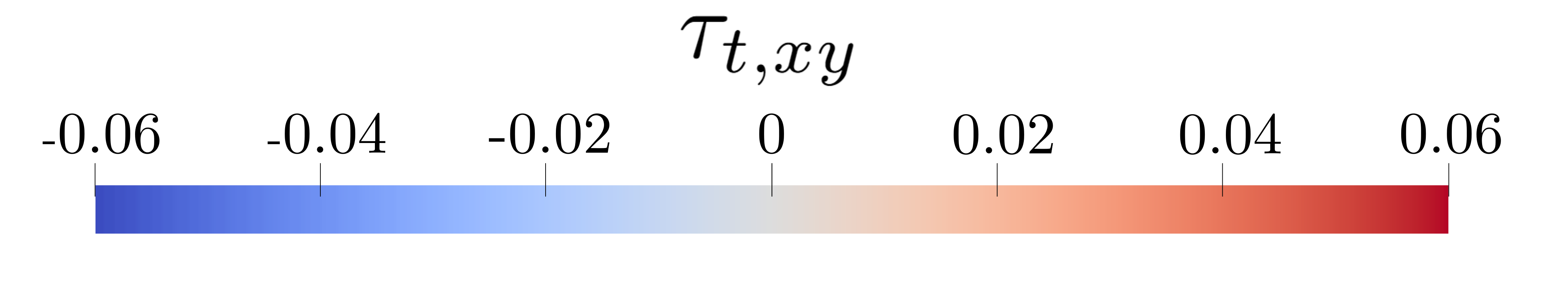}
    \caption{Shear component $\tau_{t,xy}$ of the Reynolds stress tensor. Results are shown for (a) run DDES-I0-G1, (b) the experimental data, (c) run DDES-I1-G1 and (d) run DDES-DA-G1.}
    \label{fig:ShearStressDA}
\end{figure}

At last, second-order statistics are analyzed. The shear component of the Reynolds stress tensor $\tau_{t,xy}$ is shown in \autoref{fig:ShearStressDA}. As for the pressure field, experimental results for this quantity are available but they were not used in the optimization process. One can see that none of the simulations is able to capture the high intensity turbulent shear close to the leading edge which is observed in the experiments. This problematic aspect, which is tied to the mesh resolution and the interaction of the discretization error with the turbulence model, is probably among the governing elements in the difficulty observed to match experimental data. The worst numerical results for this quantity are obtained for the simulation DDES-I0-G1, which globally produces very little turbulent shear in the separation region of the leading edge. This is probably due to the features of the inlet used. A confirmation can be found observing the profiles for the simulation DDES-I1-G1 in \autoref{fig:ShearStressDA} (c). In this case $\tau_{t,xy}$ is also underpredicted in the proximity of the leading edge, but it rapidly increases and largely exceeds the experimental prediction at around one-third of the length. Again, the DA results in \autoref{fig:ShearStressDA} (d) appear to be the closest to the experiments. Despite the underprediction at the leading edge, the distribution and intensity of $\tau_{t,xy}$ are very similar to the experimental reference. Overall, DA results exhibit a significant improvement in accuracy for all the physical quantities investigated, when compared with the preliminary DDES runs.

\section{Conclusion}
\label{sec:conclusions}

The predictive capabilities of the hybrid CFD method Delayed Detached Eddy Simulation, here used for the simulation of the BARC test case, have been improved via a data-driven technique relying on Data Assimilation. More precisely, the parametric description of a synthetic turbulent inlet condition has been optimized to reduce the discrepancy with available experimental data. A total of eight parameters have been updated, six of them representing the components of the Reynolds stress tensor while the other two are related to physical / numerical length scales. The EnKF algorithm has been adapted to bridge the available experimental observation, which consists of time-averaged features of the velocity field, and the model prediction, whose fields are intrinsically unstationary. The results obtained show that the optimization of the inlet improves the accuracy of the quantities observed in the DA process, such as the velocity field, but also of the other variables, such as the pressure coefficient at the wall and the shear component of the Reynolds stress tensor. All of these results suggest that EnKF approaches can obtain a more realistic representation of the boundary conditions in CFD problems. This is one of the main issues in the representation of complex applications such as urban flows, where intense flow accelerations over relatively short time scales can be observed and these effects can not be mitigated. The main constraints that affect the methodology used in the present work are associated with the limits in the optimization process. In fact, the number of parameters that can be manipulated is tied with the number of ensemble members available. Optimization over a very large parametric space ($\mathcal{O}(10^3-10^4)$) would require prohibitive computational resources, in terms of size of the ensemble. In addition, optimization of the functional form of modeling, in terms of structure of prescribed boundary conditions or turbulence closures, can also be performed but it is computationally expensive. Advancement on this topic is currently investigated by the team relaying on multilevel / multifidelity approaches \cite{moldovan2022multigrid,moldovan_optimized_2022}, penalization of the parametric space \cite{hou_penalized_2021} and joint application of EnKF with machine learning \cite{brajard_DA_2021}.

\section*{Acknowledgements}

The present research work has been developed in the framework of the project ANR JCJC 2021 IWP-IBM-DA. Computational resources from the cluster CASSIOPEE have been used for the calculations. The team would like to acknowledge the help from Yann Haffner from CSTB who provided the experimental data used for Data Assimilation and validation of the results. 


\bibliographystyle{elsarticle-num-names} 
\bibliography{cas-refs}

\end{document}